\documentclass[prr,twocolumn,groupedaddress,amsmath,amssymceaseaeaeab,amsfonts]{revtex4-2}
\usepackage{color,graphicx}
\usepackage{bm} 
\usepackage{booktabs}
\usepackage{subfigure}
\usepackage{mathrsf
s}
\usepackage{array}
\usepackage{braket}
\usepackage{caption}
\usepackage{subfigure}
 
\usepackage[colorlinks=true, bookmarks=true,
bookmarksnumbered=true, bookmarkstype=toc, linkcolor=magenta,
urlcolor=cyan, citecolor=cyan]{hyperref}
\begin{document}
\title{Effectiveness of quantum annealing for continuous-variable optimization}
\author{Shunta Arai$^{1}$}
\email[]{arai.s.ao@m.titech.ac.jp}
\author{Hiroki Oshiyama$^2$}
\author{Hidetoshi Nishimori$^{1,2,3}$}
\affiliation{$^1$International Research Frontiers Initiative, Tokyo Institute of Technology,
Shibaura, Minato-ku, Tokyo 108-0023, Japan\\
$^2$ Graduate School of Information Sciences, Tohoku University, Sendai 980-8579,
Japan\\
$^3$ RIKEN, Interdisciplinary Theoretical and Mathematical Sciences (iTHEMS),
Wako, Saitama 351-0198, Japan}
\date{\today}
\begin{abstract}
The application of quantum annealing to the optimization of continuous-variable functions is a relatively unexplored area of research. We test the performance of quantum annealing applied to a one-dimensional continuous-variable function with a rugged energy landscape. After domain-wall encoding to map a continuous variable to discrete Ising variables, we first benchmark the performance of the real hardware, the D-Wave 2000Q, against several state-of-the-art classical optimization algorithms designed for continuous-variable problems to find that the D-Wave 2000Q matches the classical algorithms in a limited domain of computation time. 
Beyond this domain, classical global optimization algorithms outperform the quantum device.  Next, we examine several optimization algorithms that are applicable to the Ising formulation of the problem, such as the TEBD (time-evolving block decimation) to simulate ideal coherent quantum annealing, simulated annealing, simulated quantum annealing, and spin-vector Monte Carlo. 
The data show that TEBD's coherent quantum annealing achieves far better results than the other approaches, demonstrating the effectiveness of coherent tunneling. From these two types of benchmarks, we conclude that the hardware realization of quantum annealing has the potential to significantly outperform the best classical algorithms if thermal noise and other imperfections are sufficiently suppressed and the device operates coherently, as demonstrated in recent short-time quantum simulations.
\end{abstract}
\maketitle

\section{\label{sec:sec1}Introduction}
Quantum annealing (QA) is a metaheuristic designed to solve combinatorial optimization problems by controlling quantum fluctuations \cite{Kadowaki_1998,Farhi_2001,Santoro_2002,Santoro_2006,Das_2008,Morita_2008,Tanaka_book2017,Albash_2018,Hauke_2020,Grant2020,Crosson_2021}.
The hardware implementation of QA is provided by D-Wave Quantum Inc. \cite{Dwave2010a,Dwave2010b,Dwave2010c,Dickson_2013,Lanting_2014} with more than 5000 qubits in the latest version \cite{McGeoch2022}. 

Combinatorial optimization problems with discrete variables are the main target of QA, but the basic idea of using quantum fluctuations to search for the lowest-energy state in the rugged energy landscape is also applicable to problems with continuous variables. An example is the early paper by Finnilla \textit{et al.} \cite{Finnila1994} although they used an imaginary-time version of the Schr\"odinger equation without quantum coherence.
Stella \textit{et al.} \cite{Stella_2005a} studied a few one-dimensional problems with and without rugged energy landscapes and compared QA with its classical counterpart of simulated annealing (SA) by directly solving the Schr\"{o}dinger equation for QA and the Fokker-Planck equation for SA. 

They found that the relative efficiency of QA and SA depends on the shape of the energy landscape. Simulated annealing is influenced by the height of the energy barrier, whereas QA reflects the structure of the energy spectrum, which affects the probability of quantum tunneling.
An apparent difference in performance between SA and QA was clearly seen in a parabolic washboard-like potential with a single global minimum and many local minima as proposed by Shinomoto and Kabashima \cite{Shinomoto_1991}.
These latter authors used a coarse-grained phenomenological version of the problem to show that the best possible performance of SA can be achieved when the rate of temperature decrease is proportional to the inverse of logarithm, which results in a very slow logarithmic decrease of the residual energy as a function of the annealing time. Stella \textit{et al.} discussed also QA  phenomenologically for the same washboard-like potential by considering a coarse-grained tight-binding Hamiltonian under a semi-classical approximation, which consists of a harmonic potential and a Laplacian term with a coefficient representing a semi-classical WKB tunneling effect. They found that the residual energy shows a power law decrease, which is faster than the logarithmic decrease in SA as found by Shinomoto and Kabashima.
Inack and Pilati studied the performance of QA by the projective quantum Monte Carlo method \cite{Inack_2015} and found that this imaginary-time formulation behaves better than SA in the long-time regime.
Koh and Nishimori recently analyzed a problem with a Shinomoto-Kabashima-like potential with many local minima \cite{Koh_2022}, which is known as the Rastrigin function in the field of nonconvex continuous optimization problems \cite{Dieterich_2012}.
They solved the time-dependent Schr\"{o}dinger equation directly numerically for QA and the master equation for SA and confirmed that the residual energy under QA decreases in a power law with an arbitrarily large power achievable by adjusting the annealing schedule, whereas SA has a smaller power. 
This is a simple but nontrivial example of a continuous optimization problem, in which QA has an advantage over SA.

The studies mentioned above were based on theoretical or numerical approaches.
Several experimental investigations on real hardware have recently been reported using domain wall encoding \cite{Chancellor_2019} to represent continuous variables in terms of Ising spins. Examples include Abel \textit{et al.} \cite{Abel_2021a, Abel_2021b} who observed quantum tunneling behavior from a local minimum to the global minimum in a double-well potential.  In another paper, Abel \textit{et al.} \cite{Abel_2022} have applied the same method to two-dimensional continuous optimization problems with many local minima using reverse annealing (RA) \cite{Perdomo_2011, Ohkuwa_2018, Yamashiro2019,Arai_2021}, which is a variant of QA starting from a classical spin configuration \footnote{In Ref.~\cite{Abel_2022}, RA is referred to as QA.}. 
They graphically compared the output of QA on the D-Wave hardware with data from classical continuous-variable optimization algorithms and found that the former reached the global minimum much more reliably than the latter. 
This is a very interesting development, but detailed analysis is still needed to quantitatively demonstrate a clear advantage of QA over state-of-the-art classical algorithms tailored for continuous-variable optimization.

In the present paper, we systematically investigate whether QA has an advantage over classical algorithms for optimization of the Rastrigin function in one dimension, adopting domain-wall encoding to map a continuous variable to Ising spins. The first part of our test compares the hardware D-Wave 2000Q with several well-established classical algorithms for continuous-variable optimization, Nelder-Mead (NM) \cite{Nelder_1965}, conjugate gradient descent (CGD) \cite{Hestenes_1952}, basin-hopping (BH) \cite{Leary_2000}, and differential evolution (DE) \cite{Storn_1997}.  
We have used the D-Wave 2000Q, not the latest D-Wave Advantage, because the former produced better results than the latter as will be described later, in particular in Appendix \ref{appendix_diff_annealer}. The second part also runs the hardware, but we additionally test a classical algorithm to faithfully simulate quantum annealing, TEBD (time-evolving block decimation) \cite{Vidal_2004, White_2004} based on a matrix-product representation of the wave function, in order to clarify to what extent the hardware runs coherently.  We also study several classical algorithms for combinatorial optimization, simulated annealing (SA) \cite{Kirkpatrick_1983}, simulated quantum annealing (SQA) \cite{Suzuki_1979}, and spin-vector Monte Carlo (SVMC)  \cite{Shin_2014}, to investigate how classical algorithms are affected by the barrier height compared to the quantum case.

This paper is organized as follows.
In Sec.~\ref{sec:sec2}, we explain the Rastrigin function and how to implement continuous-variable optimization with domain wall encoding.  In Sec.~\ref{sec:sec3}, we perform benchmark tests for the Rastrigin function on the D-Wave 2000Q and several classical algorithms developed for continuous-variable optimization.
In Sec.~\ref{sec::sec4}, we next compare D-Wave 2000Q, TEBD, SA, SQA, and SVMC, all of which are for discrete variable problems, to see how thermal effects influence the performance.
Finally, Sec.~\ref{sec:sec5} discusses the results and concludes the paper. A few technical details are described in the Appendixes.

\section{\label{sec:sec2}The Problem and Methods}
In this section, we explain the mathematical definition of the Rastrigin function and how to optimize the continuous function with the domain wall encoding method on the D-Wave 2000Q.
\begin{figure}[t]
\centering
\includegraphics[width=70mm]{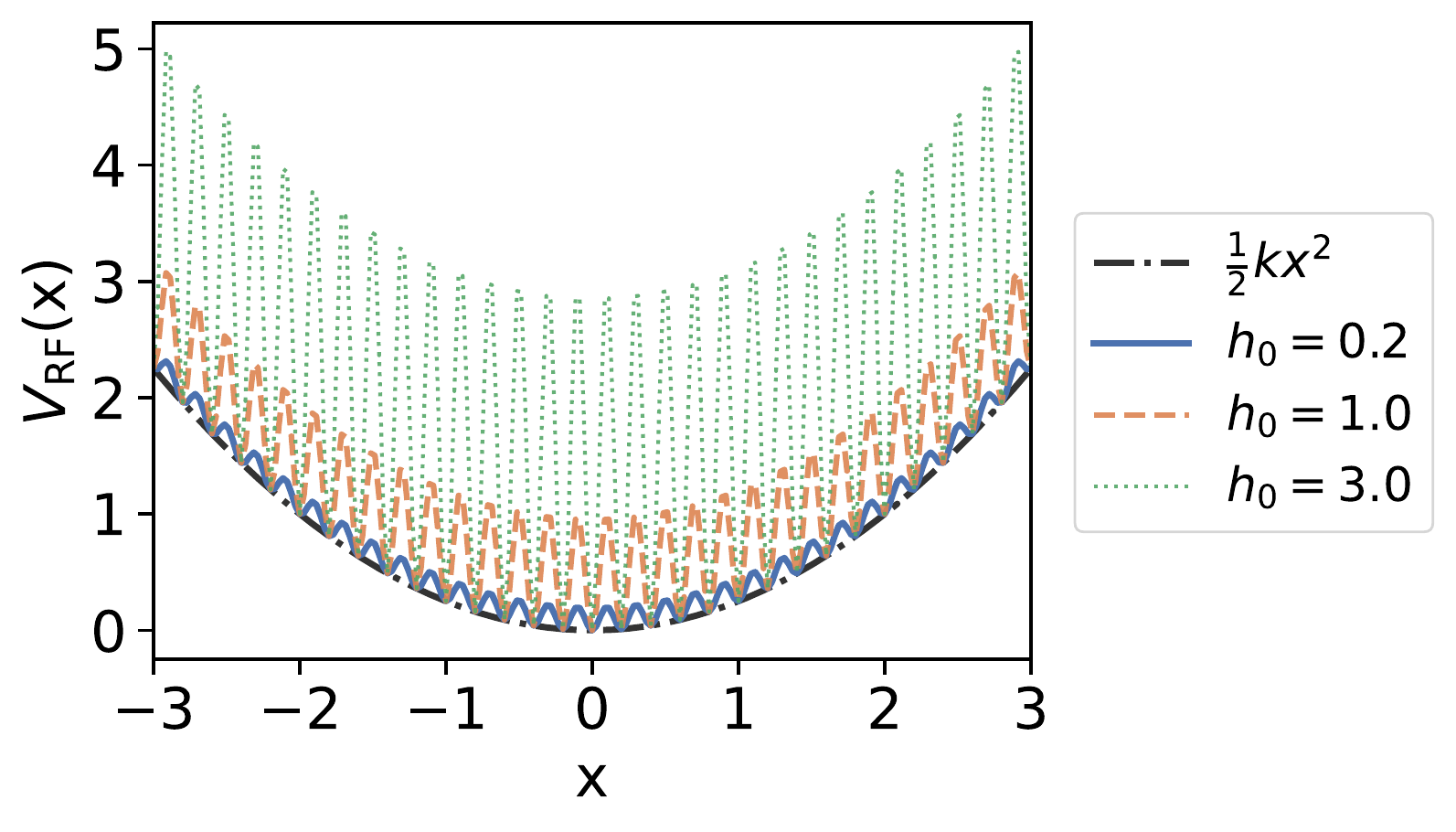}
\caption{Illustration of the Rastrigin function with $k=0.5$ and $w_0=0.2$ for different values of $h_0$. 
}
\label{fig:fig1}
\end{figure}

\subsection{\label{sec::sec21}
Rastrigin function}
The Rastrigin function is one of the standard benchmark functions for nonconvex optimization problems \cite{Dieterich_2012}.
We consider the one-dimensional Rastrigin function which consists of  a harmonic potential term with a unique global minimum at $x=0$ and a cosine term that gives many local minima as 
\begin{align}
V_{\mathrm{RF}}(x)=\frac{1}{2}kx^2+\frac{h_0}{2}\left[1-\cos\left(\frac{2\pi x}{w_0}\right)\right]
\label{eq1},
\end{align}
where $k$ is the spring constant, and $h_0$ and $w_0$ are the control parameters of the height and width of each local minimum.
The above mathematical formulation of the function was introduced by Koh and Nishimori \cite{Koh_2022} and they called it the Shinomoto-Kabashima-like potential.
The original formulation of the Rastrigin function has no control parameters and corresponds to 
$k=2$, $w_0=1$, and $h_0=20$ in Eq.~\eqref{eq1}.
In this study, we do not adopt the original formulation, but Koh and Nishimori's formulation to examine the dependence of algorithm performance on the shape of the energy landscape.
We illustrate the one-dimensional Rastrigin function with $k=0.5$ and $w_0=0.2$ for different values of $h_0$ in Fig. \ref{fig:fig1}. Due to the second cosine term, the simple gradient descent algorithm stops at a local minimum and fails to find the global minimum efficiently. 

\subsection{\label{sec::sec22}Domain wall encoding for continuous function optimization}
We briefly formulate QA and then describe the way to perform continuous function optimization by quantum annealing with domain wall encoding.

The generic Hamiltonian to be implemented by QA is
\begin{align}
H(s)&=\frac{B(s)}{2}H_0-\frac{A(s)}{2}H_{\mathrm{TF}},\label{eq2}\\
H_0&=\sum_{(i,j)\in E} J_{ij}\hat{\sigma}_i^z\hat{\sigma}_j^z+\sum_{i=1}^N h_i\hat{\sigma}_i^z
\label{eq3},\\
H_{\mathrm{TF}}&=\sum_{i=1}^N \hat{\sigma}_i^x
\label{eq4},
\end{align} where $\hat{\sigma}_i^z$ and $\hat{\sigma}_i^x$ are the Pauli operators acting at site $i$, $J_{ij}$ is the coupling constant on the edge $(i,j)$ in the graph $E$ describing the connectivity between qubits, $h_i$ is the local field at site $i$, and $N$ represents the number of spins (qubits). 
The coefficients of the transverse field and the problem are denoted as $A(s)$ and $B(s)$, respectively, controlled by the time-dependent annealing parameter $s=t/t_a\in[0,1]$.
The initial state is the ground state of Eq.~\eqref{eq4} with the coefficient $-A(0)/2$, and the system evolves from $t=0$ to $t=t_a$ following the Schr\"{o}dinger equation with the coefficients evolving as  $A(0)>0\rightarrow A(1)=0$  and 
 $B(0)\approx0\rightarrow B(1)>0$. 
 The precise forms of the scheduling functions of $A(s)$ and $B(s)$ depend on the physical devices \cite{dwave2022}.
 If the system evolves adiabatically, the final state of Eq.~\eqref{eq2} approaches the ground state of Eq.~\eqref{eq3}.

Following Refs.~\cite{Chancellor_2019,Abel_2021a, Abel_2021b},
we use the domain-wall encoding method to rewrite a continuous variable optimization problem to a corresponding problem with Ising spins. In this method, we map the value of a continuous variable to the position of a domain wall within a spin chain. Concretely, we discretize a continuous variable $x\in [x_{\mathrm{min}},x_{\mathrm{max}}]$ into $N-1$ values as
\begin{align}
x_j&~=x_{\mathrm{min}}+(j-1)\Delta x~~(j=1,\dots, N-1)
\label{eq6}
\end{align}
with the step size
\begin{align}
\Delta x&~=\frac{x_{\mathrm{max}}-x_{\mathrm{min}}}{N-1}\label{eq7},
\end{align}
and map each value $x_j$ to an $N$-spin state $\ket{x_j}$ which has a domain wall at the $j$th bond: All spins at sites 1 to $j$ are down, and the spins from $j+1$ to $N$ are up.
Using the fact that $\bra{x_j}(\hat{\sigma}_{i+1}-\hat{\sigma}_{i})/2\ket{x_j}=\delta_{ij}$ under this spin configuration, where $\delta_{ij}$ is Kronecker delta, we can construct an $N$-spin Hamiltonian $H_{\mathrm{RF}}$ such that its diagonal element in the basis $\ket{x_j}$ corresponds to $V_{\mathrm{RF}}(x_j)$ as
\begin{align}
H_{\mathrm{RF}}=\frac{1}{2}\sum_{i=1}^{N-1}V_{\mathrm{RF}}(x_i)(\hat{\sigma}_{i+1}^z-\hat{\sigma}_i^z).
\label{eq8}
\end{align}

To enforce that the ground state is a single domain wall state, we use a penalty Hamiltonian that punishes multi-kink and no-kink states as
\begin{align}
H_{\mathrm{DW}}=-J\sum_{i=1}^{N-1}\hat{\sigma}_i^z\hat{\sigma}_{i+1}^z+h\left(\hat{\sigma}_1^z-\hat{\sigma}_N^z\right),\label{eq5}
\end{align}
where the first term with $J>0$ is the ferromagnetic coupling to align neighboring spins and the second term with $h>J$ is the magnetic field at both ends of the chain to fix the spin at the first site in the down state and the spin at the $N$th site in the up state. The degenerate ground states of $H_{\mathrm{DW}}$ are $\ket{x_j}$ for $j=1,\dots,~N-1$ and the penalty energy of each additional kink is $2J$.

Combining these two Hamiltonians, the domain wall encoding Hamiltonian for optimizing $V_{\mathrm{RF}}(x)$ is given as
\begin{align}
H_0&=\lambda H_{\mathrm{RF}}+H_{\mathrm{DW}},\label{eq11}
\end{align}
where $\lambda$ is a parameter that controls the relative strength of the penalty term $H_{\mathrm{DW}}$. 
For sufficiently small $\Delta x$ and $\lambda$, the ground state of $H_0$ corresponds to the global minimum of $V_{\mathrm{RF}}(x)$.
If $\lambda$ is too large, the ground state of $H_0$ is no longer a single domain wall state.

Assuming that the potential function is differentiable and $\Delta x \ll 1$, Eq.~\eqref{eq8} can be reduced to 
\begin{align}
H_{\mathrm{RF}}&=-\frac{1}{2}\Bigg(\Delta x\sum_{i=2}^{N-1}\frac{V_{\mathrm{RF}}(x_i)-V_{\mathrm{RF}}(x_{i-1})}{\Delta x}\hat{\sigma}_i^z\nonumber\\
&\quad +V_{\mathrm{RF}}(x_{1})\hat{\sigma}_1^z-V_{\mathrm{RF}}(x_{N-1})\hat{\sigma}_N^z\Bigg)\nonumber\\
&\approx-\frac{1}{2}\Bigg(\Delta x\sum_{i=2}^{N-1}\left.\frac{d V_{\mathrm{RF}}(x)}{d x}\right|_{x=x_i}\hat{\sigma}_i^z\nonumber\\
&+V_{\mathrm{RF}}(x_{1})\hat{\sigma}_1^z-V_{\mathrm{RF}}(x_{N-1})\hat{\sigma}_N^z\Bigg)\nonumber\\
&=\sum_{i=1}^{N}h_i\hat{\sigma}_i^z,\label{eq9}\\
h_i&=\begin{cases}
-\frac{V_{\mathrm{RF}}(x_{1})}{2}\qquad (i=1)\\
-\frac{\Delta x}{2}\left.\frac{d V_{\mathrm{RF}}(x)}{d x}\right|_{x=x_i}\,(i=2,...,N-1)\\
\frac{V_{\mathrm{RF}}(x_{N-1})}{2} \qquad (i=N)
\end{cases}
\label{eq10}.
\end{align}

The above construction of the Ising Hamiltonian for continuous function optimization can be easily generalized to higher-dimensional problems \cite{Abel_2021b}.

\section{\label{sec:sec3}Comparison of QA with classical algorithms for continuous optimization}
In this section, we show the results of benchmarks for QA and several classical algorithms designed for the optimization of continuous variable functions to investigate how QA applied to continuous variable optimization compares with dedicated classical algorithms.  In the present section, QA means the hardware implementation, the D-Wave 2000Q.

\subsection{\label{sec::sec31}Optimization algorithms}
We run QA and the four classical algorithms for the optimization of continuous-variable functions: Nelder-Mead (NM) \cite{Nelder_1965}, conjugate gradient descent (CGD) \cite{Hestenes_1952}, basin hopping (BH) \cite{Leary_2000}, and differential evolution (DE) \cite{Storn_1997}, which are easily performed on the classical computer and are commonly used for the optimization of continuous functions with a complex potential landscape.
While the NM and CGD methods are for local optimization, BH and DE are for global optimization. 
The NM method searches for minimum by updating the polyhedron with the observed functional value.
The CGD method utilizes the gradient of the potential function with the adjustable step size to find the minimum.
Basin hopping performs two steps iteratively: local optimization around the proposed point and perturbation to search for the neighborhood domain of the current solution.
Differential evolution is a kind of evolutionary computation and updates a proposed point by following the evolutionary procedure.
Details of the classical algorithms are explained in Appendix \ref{appendix_a}.  Domain-wall encoding is applied only to QA on the D-Wave 2000Q, and classical algorithms use the continuous variable directly.

We apply these algorithms to Eq.~\eqref{eq1} using the Scipy package \cite{Scipy2020}.
All numerical simulations for the classical optimization algorithms are performed on a laptop with Apple M1 Max chip with 10 CPU cores, which is one of the fastest chips \footnote{Extensive CPU benchmarks are performed in Geekbench \cite{geek}.
 The single-core score of the Apple M1 Max chip in Geekbench ver.~6 is higher than those of the Intel Xeon E5-1650 used in an old benchmark \cite{Denchev_2016} and the Intel Xeon E5-2695 v4 in a recent benchmark \cite{Tasseff_2022}. }.
For QA, we optimize Eq.~\eqref{eq11} with the  D-Wave 2000Q, more precisely the \textrm{``DW\_2000Q\_6"} device.
We have also run the same experiments on two types of the more recent D-Wave Advantage. Since \textrm{``DW\_2000Q\_6"} yielded the best results, we utilize it in the main text. 
The comparison of performance among a few models of D-Wave devices is described in the Appendix \ref{appendix_diff_annealer}.

\subsection{\label{sec::sec32}Measured quantities} 
We compute two observables. 
The first is the ground state probability,
\begin{align}
P_{\mathrm{GS}}=\frac{\rm\#GS}{n_{\mathrm{init}}}, 
\label{pgs}
\end{align}
where $\rm\#GS$ is the number of cases where the global minimum was obtained by each algorithm, and $n_{\mathrm{init}}$ stands for the number of trials (the number of different initial conditions). 
The second is the absolute error in the energy,
\begin{align}
E_{\mathrm{abs}}=\big|\langle V_{\mathrm{RF}}(x)\rangle - V_0\big|,
\end{align}
where the brackets represent the average over the different initial conditions and $V_0=0$ is the minimum value of Eq.~\eqref{eq1}.

\subsection{\label{sec::sec33}Parameter settings}
For the classical optimization algorithms, we set $n_{\mathrm{init}}=10^3$ and fix the search space as $x\in [-3,3]$ \footnote{In the original formulation of the Rastrigin function, the search space is defined in  $x\in [-5.12,5.12]$ \cite{Dieterich_2012}.
Taking into account the computational precision of QA,
we reduce the search space
}.
For a fair comparison with QA, 
 we regard the output $x$ obtained by the classical optimization algorithms as the solution to Eq.~\eqref{eq1} if the condition $|x|<\Delta x$ is satisfied.
 We control the execution time $t_{\mathrm{ex}}$  by changing the maximum number of iterations of classical algorithms as $t_{\mathrm{max}}=2^i$ with the integer $i$ running from 0 to 7.
 The execution time can be computed as 
$t_{\mathrm{ex}}=\sum_{k=1}^{t_{\mathrm{max}}}\tau_k$ where $\tau_k$ represents the $k$th iteration time.

For the NM and CGD methods, the default parameter settings in Scipy are used.
For BH, we apply the CGD method in the local optimization process.
For DE, the number of populations is set to 10 and 120.
Each algorithm stops on the way if it satisfies the stopping criterion as $|x^{t}-x^{t-1}|<10^{-7} \quad (t=1,\dots,t_{\mathrm{max}})$.

Our experimental settings are carefully designed for a fair comparison of QA and the classical algorithms in execution time and performance.
Although there may be room for further improvements in execution time and performance, our studies are expected to lead to a general understanding of the features of QA for continuous function optimization against classical optimization algorithms.

For QA, we set the system size $N=211$ and fix the coefficients as $J=1,h=2$, and $\lambda=1$.  The value $N=211$, which corresponds to $\Delta x=0.0286$ with $-x_{\rm min}=x_{\rm max}=3$, is large enough to represent the Rastrigin function with sufficient precision in its hills and valleys.
The gradient of the potential in Eq.~\eqref{eq1} is equal to $0$ at the global minimum $x=0$ and yields the vanishing of the magnetic field $h_{j'}=0$ where $j'$ represents the place of the zero magnetic field. 
The vanishing of the magnetic field according to Eq.~\eqref{eq10} gives two degenerate ground states of Eq.~\eqref{eq11} as $\bm{\hat{\sigma}^z}_{\mathrm{\pm}}=(-1,\dots,-1,\hat{\sigma}_{j'}^z=\pm1,1,\dots,1)$.
The ground state $\bm{\hat{\sigma}^z}_{\mathrm{-}}$ yields $x_{j'}=0$ with Eq.~\eqref{eq6}.
On the other hand, the global minimum of Eq.~\eqref{eq1} can not be recovered from the ground state $\bm{\hat{\sigma}^z}_{\mathrm{+}}$ \footnote{By substituting $x_j=0$ into the right-hand of Eq.~\eqref{eq6}, the location of the zero magnetic field can be computed as $j'=106$ with the present numerical values, $N=211$ and $x_{\rm max}=-x_{\rm min}=3$.
For $\hat{\sigma}_{106}^{z}=-1$, the ground state $x=0$ can be recovered as $x_{106}=0$.
For $\hat{\sigma}_{106}^{z}=+1$, the domain wall is located between $j=105$ and $j=106$, and $x_{105}=-\Delta x(\ne 0)$ results
}.
Since the problem is negligible when $N$ is large enough, we regard the two degenerate ground states of Eq.~\eqref{eq11} as the equally qualified solution of Eq.~\eqref{eq1}.

We operate the execution time of QA by varying the annealing time $1 (\mu s) \leq t_a\leq 500 (\mu s)$.
The total execution time of QA
is defined as $t_{\mathrm{ex}}=t_{p}+t_{s}$ where $t_{p}$ is programming time and $t_{s}$ is sampling time which includes annealing time and readout time. 
The $t_{\mathrm{ex}}$ can be obtained from \textit{"qpu\_access\_time"} in the D-Wave Ocean SDK 
 \cite{dwave2022}.
To compute the observables, 
we repeat the annealing process 20 times independently and sample $n_{\mathrm{reads}}=10^3$ spin configurations for each run.
To reduce control errors, we apply flux bias calibration and gauge transformation \cite{Nishimura_2021, King_2022}.
The flux bias is estimated before each run.
A gauge is generated from a uniform distribution and fixed in each independent run. 
The ground-state probability is calculated from $n_{\mathrm{reads}}=10^3$ spin configurations. 
The absolute error in the energy is computed from samples that meet the single-domain-wall condition.

\subsection{\label{sec::sec34}Result for high energy barrier}
We first consider the problem with a high energy barrier $h_0=1.0$.
We show the dependence of the ground state probability and the absolute error in the energy
on the execution time in Fig.~\ref{fig:fig_2}. 
Data points denoted as ``D-Wave (mean)" are the averaged result for $20$ independent runs. 
Data with the maximum ground state probability among all runs are denoted as ``D-Wave (best)".  
The error bars show the standard error computed by bootstrapping.
Execution time $t_{\mathrm{ex}}$ is always expressed in seconds.
 
Figure \ref{fig:fig_2} (a) shows that around the region $0.7 ({\rm sec}) \leq t_{\mathrm{ex}} \leq  0.9 ({\rm sec})$, the ground state probability obtained by ``D-Wave (best)" can be larger than the other algorithms except for ``DE (popsize=120)".
 The data for the NM and CGD methods are not shown in the left-side range beyond each left-most point because they failed to find the solution in this short-time region.
Also, the data for the NM and CGD methods as well as for QA are not shown to the right of their respective rightmost points since they show saturation and no improvements in these long-time region \footnote{The NM and CGD methods get stuck in local minima before the number of iterations of the algorithms reaches the maximum value.
Therefore, an increase in the maximum number of iterations does not affect the execution time, and their rightmost points have the same values of the execution time}.
 On the other hand,
BH and DE find the solution in the longer time region $ 10 ({\rm sec})\leq t_{\mathrm{ex}} $ and their performance improves as the execution time increases.
In the shorter time region, the minimum execution time for BH and DE 
cannot be reduced further than shown in Fig.~\ref{fig:fig_2} with our computational resources.

Figure \ref{fig:fig_2} (b) shows that the absolute error in the energy obtained by QA is smaller than that of the NM and CGD methods and BH around the region $0.4 (\rm{sec})\leq t_{\mathrm{ex}}\leq  0.9 (\rm{sec})$.
Therefore, QA can find the local minima closer to the global minima than the NM and CGD methods and BH in this region. 
The outputs of BH and DE with large populations concentrate around the global minima as we increase the execution time as we see in the right half of Fig.~\ref{fig:fig_2} (b).
If the number of populations is small, 
DE falls into local minima even in the long execution time region.
Similarly to the ground-state probability, the absolute error in the energy of QA saturates at a finite value.

For the higher energy barrier $h_0=3.0$, the qualitative behaviors of the classical algorithms are similar to those of $h_0=1.0$. 
On the other hand, the performance of QA deteriorates significantly, as will be described in
Sec.~\ref{sec::sec41}. 

\subsection{\label{sec::sec35}Result for low energy barrier}
\begin{figure}[t]
\centering
\includegraphics[width=0.8\hsize]{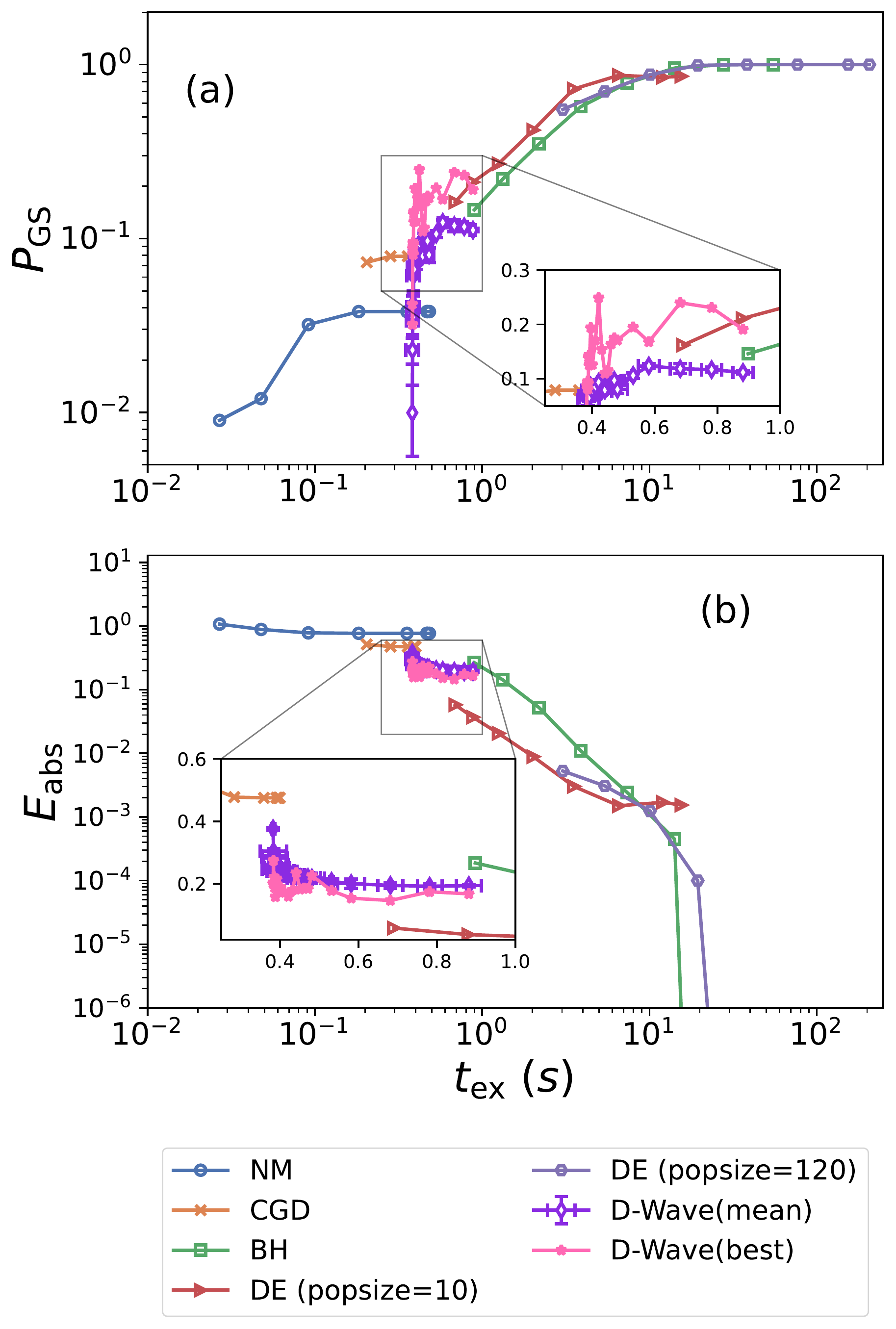}
  \caption{
 Dependence of observables on the execution time for $k=0.5$, $w_0=0.2$, and $h_0=1.0$. 
The vertical axes denote the observables: (a) the probability of ground state and (b) the absolute error in energy.
The insets enlarge the region $0.25 (\mathrm{sec})\leq t_{\mathrm{ex}}\leq 1 (\mathrm{sec})$.
  }
  \label{fig:fig_2}
\end{figure}

\begin{figure}[t]
\centering
\includegraphics[width=0.8\hsize]{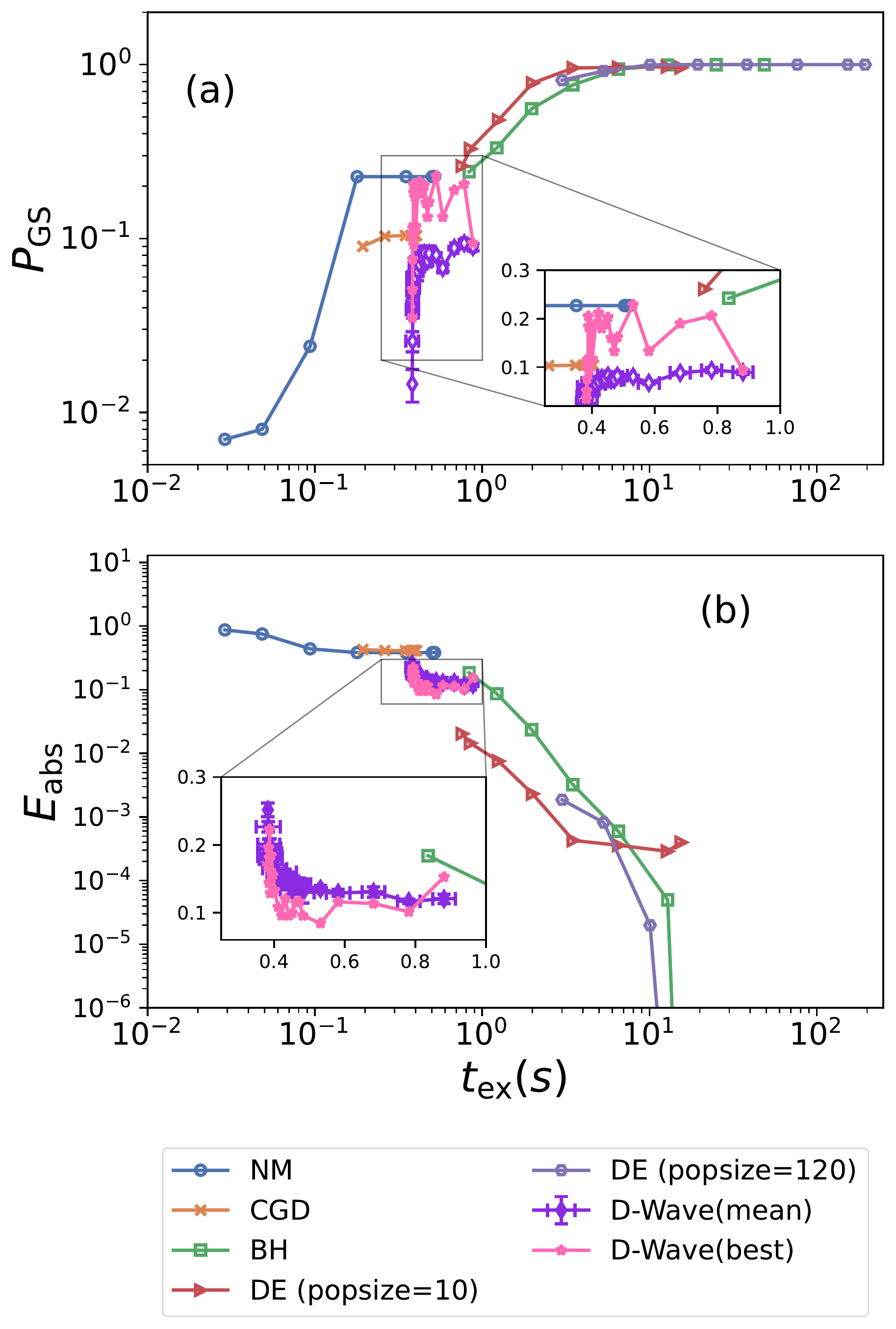}
\caption{Dependence of observables on the execution time for $k=0.5$ and $w_0=h_0=0.2$.
Both axes are the same as those in Fig. \ref{fig:fig_2}.
The insets enlarge the region $0.25 (\mathrm{sec})\leq t_{\mathrm{ex}}\leq1 (\mathrm{sec})$.
  }
  \label{fig:fig_3}
\end{figure}

Next, we consider the problem with a lower energy barrier $h_0=0.2$.
Figure \ref{fig:fig_3} (a) shows an overall improvement of classical algorithms compared to the case of a higher energy barrier in Fig.~\ref{fig:fig_2} (a).
Nevertheless, the performance of QA deteriorates slightly.
The best result from QA does not reach the level obtained by BH and DE, even with a minimum execution time of about 0.8 seconds.
The saturation of the ground-state probability of QA for a longer execution time is observed beyond about 0.9 seconds as in Fig.~\ref{fig:fig_2} (a) and is thus not drawn explicitly in Fig.~\ref{fig:fig_3} (a).

Figure \ref{fig:fig_3} (b) demonstrates that
the absolute error in the energy generally decreases compared to the case of a high-energy barrier in Fig.~\ref{fig:fig_2} (b).
The lowest value of the absolute error in the energy for the NM method is seen to be slightly lower than that of the CGD method in contrast to the case of Fig.~\ref{fig:fig_2} (b).
This result implies that the one-dimensional structure with unique global minima fits the search strategy of the NM method.
Except for this observation, the behaviors of the absolute error in the energy remain similar to those in Fig. \ref{fig:fig_2} (b).

In summary, the performance of QA is slightly better than those of local optimization algorithms and is comparable to that from global algorithms in a limited region of execution time for the potential function with a higher energy barrier.
On the other hand, if the energy barrier is low, we did not observe the clear benefits of QA over classical algorithms for the present problem of continuous function optimization.
Whereas increasing the execution time enhances the performance of the global optimization algorithms, QA shows saturation beyond intermediate time scales.
In the next section, we investigate whether this saturation originates in the problem in hardware implementation or in other reasons including the mapping of the continuous variable to Ising spins by the domain-wall encoding.
\begin{figure*}[t]
\centering
\includegraphics[width=0.7\hsize]{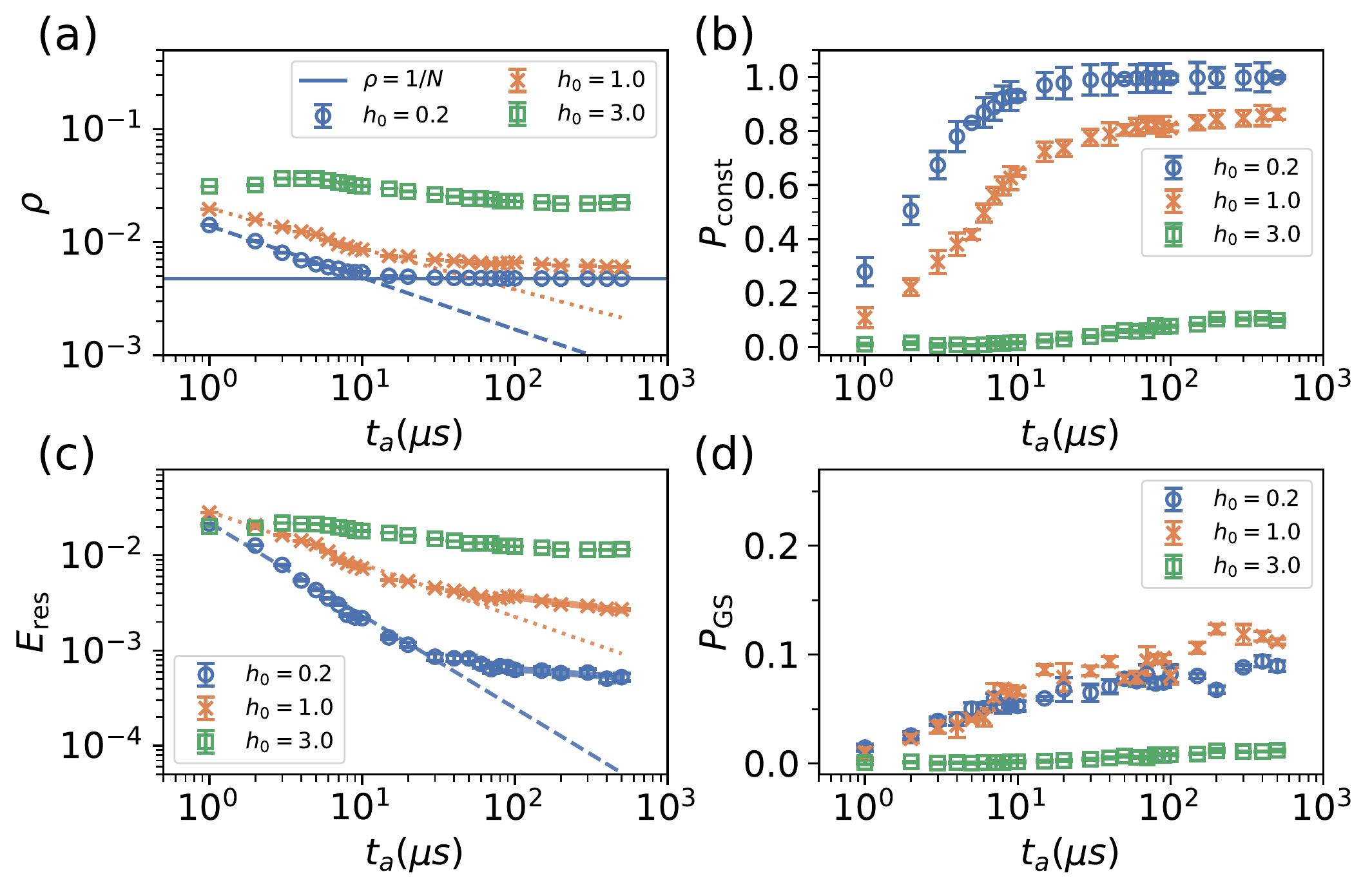}
  \caption{Dependence of the observables obtained by the D-Wave on the annealing time for three values of $h_0$.
  The vertical axes are the observables: (a) the kink density, (b) the probability satisfying the single domain wall condition, (c) the residual energy, and (d) the ground state probability.
  The horizontal axes represent the annealing time.
  }
  \label{fig:fig_4}
\end{figure*}
\section{\label{sec::sec4}Comparison of QA with algorithms for discrete-variable optimization}

In this section, we compare the performance of the D-Wave 2000Q with several classical discrete-variable optimization algorithms, as well as with TEBD, a direct implementation of noise-free coherent QA on a classical computer, to examine how the D-Wave 2000Q compares with these other methods on the same footing of domain-wall coding.  The results may help us estimate how the hardware would perform if the noise level were significantly reduced. The classical algorithms used in this section are SA, SQA, and SVMC. We carefully distinguish between coherent QA simulated by TEBD and hardware implementation of QA, the latter being represented by the product name.
For the sake of simplicity, the D-Wave 2000Q is referred to as the D-Wave hereafter.
\begin{table*}[t]
  \centering
  \caption{The fitting exponents of the observables in the short annealing time region: (a) the kink density and  (b) the residual energy.}
  \scalebox{0.9}{ 
  \subtable[]{
  \begin{tabular}{|c||c|c|c|c|c|c|}  \hline
    $h_0$ & D-Wave&Coherent-QA&SA&SQA  \\ \hline 
    0.2 &  $0.458\pm0.014$ 
    &$0.517\pm0.003$
    & $0.616\pm0.006$
    &$0.631\pm0.014$
    \\ \hline
    1.0 & $0.359\pm0.011$
    &$0.529\pm0.003$
    & $0.474\pm0.017$
    &$0.571\pm0.01$
    \\ \hline
    3.0 & -
    &$0.596\pm0.008$
    & $0.164\pm0.011$
    &$0.148\pm0.026$
    \\ \hline
  \end{tabular}
  } }\\
  \scalebox{0.9}{ 
  \subtable[]{
     \begin{tabular}{|c||c|c|c|c|c|c|}  \hline
    $h_0$ & D-Wave&Coherent-QA& SA&SQA  \\ \hline 
    0.2 &  $0.974\pm0.022$ 
    &$0.543\pm0.010$
    & $0.632\pm0.012$
    &$0.651\pm0.015$\\ \hline
    1.0 & $0.553\pm0.025$ 
    &$0.601\pm0.006$
    & $0.483\pm0.014$
    &$0.591\pm0.011$
    \\ \hline
    3.0 & -
    &$0.657\pm0.002$
    & $0.182\pm0.01$
    &$0.244\pm0.015$
    
    \\ \hline
  \end{tabular}
  }}
  \label{table_1}
\end{table*}
\begin{table*}[t]
  \centering
  \caption{The fitting exponents of the residual energy in the long annealing time region.}
  \scalebox{0.9}{ 
  \begin{tabular}{|c||c|c|c|c|}  \hline
    $h_0$ &D-Wave& SA&SQA&SVMC 
    \\ \hline 
    0.2 &$0.124\pm0.03$
    &$0.307\pm0.025$ 
    &$0.66\pm0.044$
    & $1.043\pm0.059$
    \\ \hline
    1.0 &$0.201\pm0.02$
    & $0.291\pm0.031$
    &$0.786\pm0.032$
    & $1.722\pm0.169$
    \\ \hline
    3.0 & -
    & -
    &$0.751\pm0.009$
    & -
    \\ \hline
  \end{tabular}
    \label{table_2}
    }
\end{table*}

\begin{figure}[t]
\centering
\includegraphics[width=0.75\hsize]{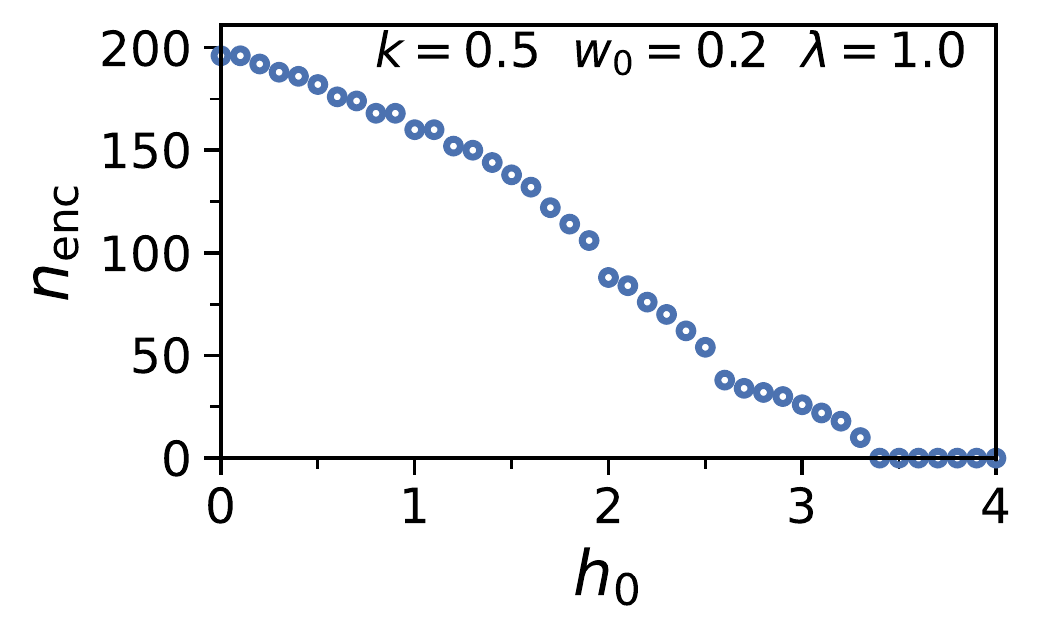}
\caption{Number of single kink states encoded in the lowest energy states of the classical Hamiltonian $H_0$ as a function of the height of energy barrier $h_0$. The parameters of the Rastrigin function are set to $k=0.5$ and $w_0=0.2$, and the weight of $H_{\mathrm{RF}}$ is set to $\lambda=1.0$.}
  \label{fig:n_enc}
\end{figure}

We compute four quantities. 
The first one is the kink density
\begin{align}
\rho=\frac{1}{2N}\sum_{i=1}^{N-1}\left(1-\langle\hat{\sigma}_i^{z}\hat{\sigma}_{i+1}^{z}\rangle\right). 
\end{align}
The second is the probability $P_{\mathrm{const}}$ that the result satisfies the condition of single domain wall, and the third one is the residual energy 
\begin{align}
E_{\mathrm{res}}=\frac{\langle H_0\rangle - E_0}{N}.
\end{align}
Here, the brackets $\langle \cdots \rangle$ denote the sample average and $E_0$ is the ground state energy of Eq.~\eqref{eq11}. The last quantity is the ground state probability $P_{\mathrm{GS}}$ that the ground state is successfully found. 
The kink density $\rho$ represents the number of defects in spin chains. 
When a single domain wall exists, $\rho=1/N$ holds.
Unlike the absolute error $E_{\mathrm{abs}}$ used in Figs.~\ref{fig:fig_2} (b) and \ref{fig:fig_3} (b), the residual energy $E_{\mathrm{res}}$ is calculated from all samples, including the samples which do not satisfy the single domain wall condition.
To understand the behavior of the entire system, we do not adopt $E_{\mathrm{abs}}$ but use $E_{\mathrm{res}}$ in this section.

Since the execution time of the D-Wave contains the measurement, initialization and other auxiliary-processing times, we mainly use the pure annealing time as the $x$ axis of each figure for fair comparison.
The experimental setting of the D-Wave is the same as those in Figs.~\ref{fig:fig_2} and \ref{fig:fig_3}.

\subsection{\label{sec::sec41}D-Wave}
Figure \ref{fig:fig_4} shows the dependence of the observables obtained from the D-Wave on the annealing time for $h_0=0.2, 1.0$ and 3.0.
The error bars represent the standard error obtained by bootstrapping and computed in $2\times 10^4$ samples for $\rho$ and $E_{\rm res}$ and in $20$ independent runs for $P_{\rm const}$ and $P_{\rm GS}$.
All error bars in figures of this section are computed in this way.
The data for $h_0=0.2$ and $h_0=1.0$ are the same as those used in Figs.~\ref{fig:fig_2} and \ref{fig:fig_3}.

\subsubsection{Kink density}
Let us first look at the kink density $\rho$ in Fig.~\ref{fig:fig_4} (a).
We see that it needs some time that the data reaches $\rho=1/N$, where the condition of single domain wall is satisfied. In other words, this is the minimum amount of time it takes to find the correct solution on average.  If the kink density behaves as $\rho\propto t_a^{-b}$ below this threshold time, the annealing time required for the kink density to reach $\rho=1/N$ scales as $t_a\propto N^{1/b}$. Therefore, the value of the exponent $b$ is important to understand how the performance of the domain wall encoding depends on $N$, and a larger $b$ is preferable.
Except for the case of $h_0=3.0$, Fig.~\ref{fig:fig_4} (a) demonstrates that the kink density gradually decreases as we increase the annealing time.

Each line is fitted with the polynomial function  $\rho \propto t_a^{-b}$ in the range $1 (\mu s) \leq t_a \leq10 (\mu s)$.  Extracted exponents are summarized in Table \ref{table_1}. 
A smaller value of the exponent reflects the effect of the higher energy barrier as seen in Table \ref{table_1} (a).
In addition, the higher energy barrier gives a larger absolute value of $\rho$, not just a smaller exponent.
For $h_0=3.0$, the kink density behaves non-monotonically. Such behavior has also been observed in several experiments on the D-Wave devices and is believed to be caused by the open-system nature of the hardware \cite{Weinberg_2020,King_2022}.

Kink densities for $h_0=1.0$ and $3.0$ do not reach $\rho=1/N$ and saturate at finite values for long annealing time.
Since the behaviors of all observables in Fig.~\ref{fig:fig_4} do not change even though we increase the annealing time beyond $t_a=500(\mu s)$, we only show the data in $1 (\mu s) \leq t_a\leq 500 (\mu s)$.

\subsubsection{Single domain wall condition}
Figure \ref{fig:fig_4} (b) illustrates that almost all samples satisfy the single domain wall condition for $h_0=0.2$ beyond $t_a=10(\mu s)$.
On the other hand, the probability of satisfying the single domain wall condition saturates around $P_{\mathrm{const}}\approx0.8$ for $h_0=1.0$ and $P_{\mathrm{const}}\approx0.1$ for $h_0=3.0$. 
 For those cases with higher energy barriers, the number of samples that satisfy the single domain wall condition deteriorates, and an excessive number of domain walls appear.

\subsubsection{Residual energy}
Figure \ref{fig:fig_4} (c) indicates that the residual energy polynomially decreases as $E_{\mathrm{res}}\propto t_a^{-c}$ in the short annealing time region $1 (\mu s) \leq t_a \leq10 (\mu s)$ except for $h_0=3.0$.
As with the kink density, the exponent of $E_{\mathrm{res}}$ becomes smaller for higher energy barriers, see Table \ref{table_1} (b). 
Higher energy barriers yield larger absolute values of the residual energy in addition to smaller exponents.

Next, we investigate the behaviors of $E_{\mathrm{res}}$ in the large annealing time region  $ t_a \geq 100 (\mu s)$. The scaling analysis in this region gives us a hint on the performance of the D-Wave as an optimization solver. 
We omit the data for $h_0=3.0$ because few samples satisfy the single domain wall condition. 
We fit the polynomial function and show the exponents in Table \ref{table_2}.
For the problem with $h_0=1.0$, the slope is slightly larger than in the case of $h_0=0.2$.
This result indicates interestingly that the process is more efficient in the problem with the higher energy barrier. 

Although the residual energy has a polynomial decay as found in Koh and Nishimori \cite{Koh_2022}, the exponent of $E_{\mathrm{res}}$ is not close their result $E_{\mathrm{res}}\sim t_a^{-2}$. This discrepancy may come from the effects of noise on the hardware.
In Fig.~\ref{fig:fig_4}, we omit the data for $t_a\geq 500 (\mu s)$ because the behavior of $E_{\mathrm{res}}$ hardly changes, and the effect of noise increases in the longer annealing time. 
This result corresponds to the fact in Figs.~\ref{fig:fig_2} and \ref{fig:fig_3} that the observables hardly show improvements if we increase the execution time beyond a certain value and the data are omitted over $t_{\mathrm{ex}}\geq 0.9 (\mathrm{sec})$.
Therefore, the execution time region $t_{\mathrm{ex}}\geq 0.9 (\mathrm{sec})$ corresponds to the annealing time region  $t_a\geq 500  (\mu s)$.

\subsubsection{Ground state probability}
The ground state probability for $h_0=1.0$ is slightly larger than for $h_0=0.2$ but remains small for $h_0=3.0$ as seen in Fig.~\ref{fig:fig_4} (d).
For $h_0=0.2$ and $1.0$, the residual energy shows slightly decreasing behavior even in the large annealing time region $t_a\geq 100(\mu s)$, but the ground state probability saturates around $P_{\mathrm{GS}}\approx0.1$.  This result indicates that the transition to the ground states hardly appears, whereas the transition to the neighborhoods of the ground states occurs
in this time scale. 
The ground state probability for $h_0=3.0$ is suppressed due to the existence of many defects with few samples satisfying the single domain wall condition as shown in Figs.~\ref{fig:fig_4} (a) and (b).

\begin{figure*}[t]
\centering
\includegraphics[width=0.7\hsize]{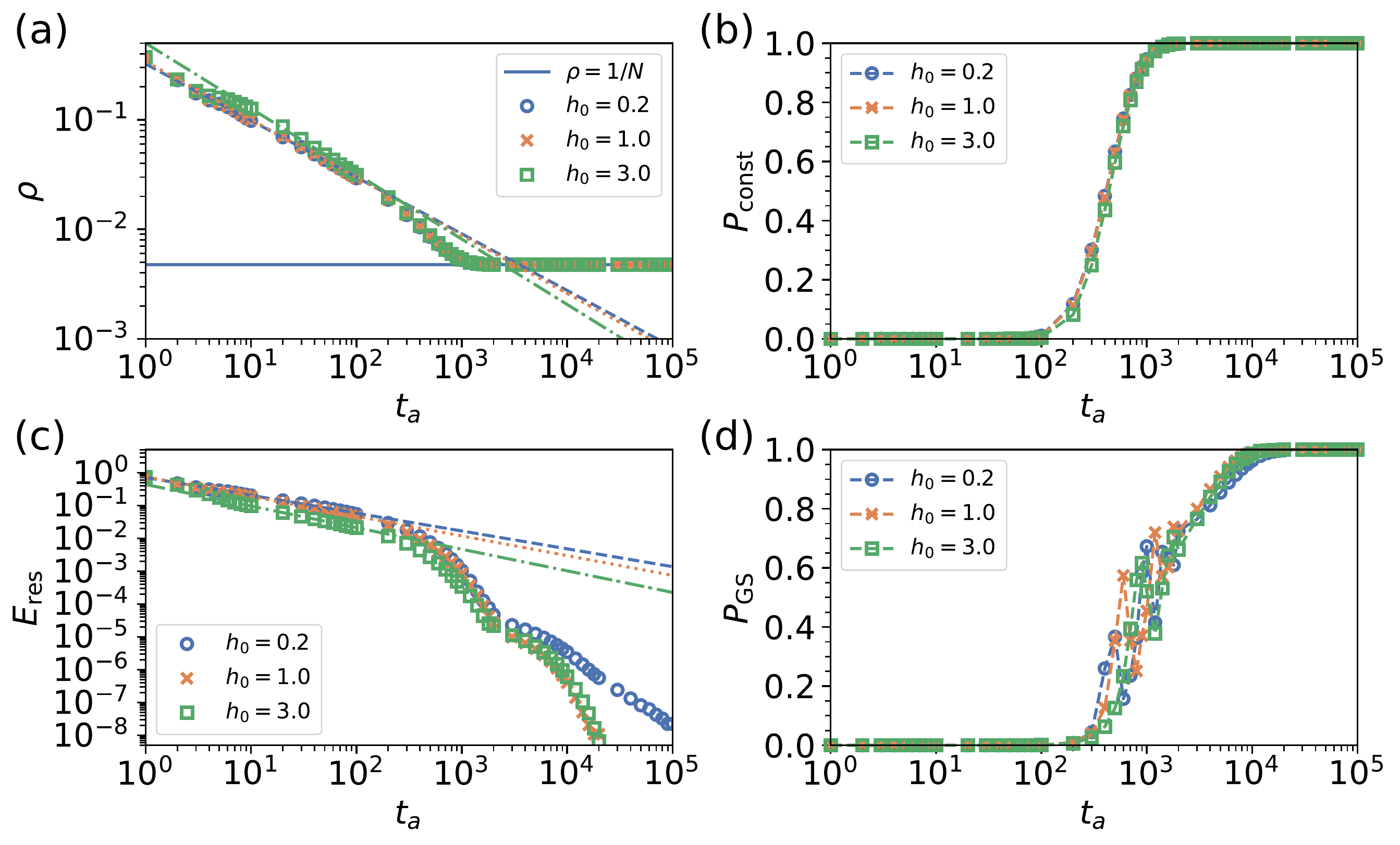}
  \caption{Results of TEBD simulations for different values of $h_0$ as functions of annealing time. 
The vertical axes are the same as those in Fig.~\ref{fig:fig_4}. The annealing time $t_a$ is in the unit of $J$ and does not correspond to that of Fig.~\ref{fig:fig_4}.
}
  \label{fig:TEBD1}
\end{figure*}
\subsubsection{Kinks in low-energy states}
To understand the reason for the significant performance deterioration observed for $h_0=3.0$, we examine the behavior of the low energy states of the classical Hamiltonian $H_0$ in Eq.~\eqref{eq11}.
To evaluate how many single-kink states are properly encoded as low energy states of $H_0$, we introduce $n_{\mathrm{enc}}$ defined as
\begin{align}
n_{\mathrm{enc}}=\min\{j\in0,1,\dots\mid n_{\mathrm{kink}}(\bm{S}_j)\neq1\},\label{eq:n_enc}
\end{align}
where $n_{\mathrm{kink}}(\bm S)$ denotes the number of kinks in the spin state $\bm S$, and $\bm{S}_j$ is the spin state corresponding to the $j$th energy level. 
The value of $n_{\mathrm{enc}}$ represents the index of the lowest energy level for which the number of kinks in the corresponding spin state is not equal to 1.
If $n_{\mathrm{enc}}>0$, the ground state is a single kink state. If $n_{\mathrm{enc}}=0$ on the other hand, the ground state is a multi-kink state, indicating a failure of the encoding. 
The value of $n_{\mathrm{enc}}$ depends on the relative magnitude of $\lambda H_{\rm RF}$ and $H_{\mathrm{DW}}$. If we set $\lambda$ to be sufficiently small relative to $J$, we can make $n_{\mathrm{enc}}>0$. 

Figure \ref{fig:n_enc} shows the $h_0$ dependence of $n_{\mathrm{enc}}$ obtained by dynamic programming \cite{Hamasaki_2004}. 
As $h_0$ increases, the values of the magnetic field $h_i$ determined by Eq. (\ref{eq12}) become larger, resulting in a smaller $n_{\mathrm{enc}}$. For $h_0=0.2$ and $1.0$, we have $n_{\mathrm{enc}}=196$ and $160$, respectively, indicating that most of the single kink states are encoded as low energy states of $H_0$. For $h_0=3.0$, we have $n_{\mathrm{enc}}=26$, indicating that the lowest energy states of $H_0$ are contaminated with multi-kink states. 
The reason for the extremely low constraint satisfaction probability $P_{\mathrm{const}}$ for $h_0=3.0$ in Fig.~\ref{fig:fig_4} (b) is considered to be that the low energy multi-kink states trap the annealing process. Reducing $\lambda$ increases $n_{\mathrm{enc}}$, which leads to an increase in $P_{\mathrm{const}}$.
However, reducing $\lambda$ also increases the thermal effect, and therefore, the optimization performance may not necessarily improve. The dependence of the observables on $\lambda$ is shown in Appendix. \ref{appendix_b2}.
\subsubsection{Summary}
The values of observables obtained by the D-Wave depend on the height of the energy barrier. 
From the viewpoint as an optimization solver, achievement of the condition $P_{\mathrm{const}}=1$ is important.  For the problem with the higher energy barrier, the probability of satisfying the single-domain wall condition does not reach 1 in our experiments even if we increase the annealing time. The ground state probability also saturates at a finite value and does improve any further. Since the decoherence effects by thermal fluctuations may be dominant in the long annealing time region, these effects hamper the transition to the ground states and degrade the performance.  Although the residual energy $E_{\mathrm{res}}$ shows a polynomial decay in this region, the exponents are different from the direct simulation of the Schr\"{o}dinger dynamics \cite{Koh_2022} due to the effects of thermal noise.
 
\begin{figure*}[t]
\centering
\includegraphics[width=0.85\hsize]{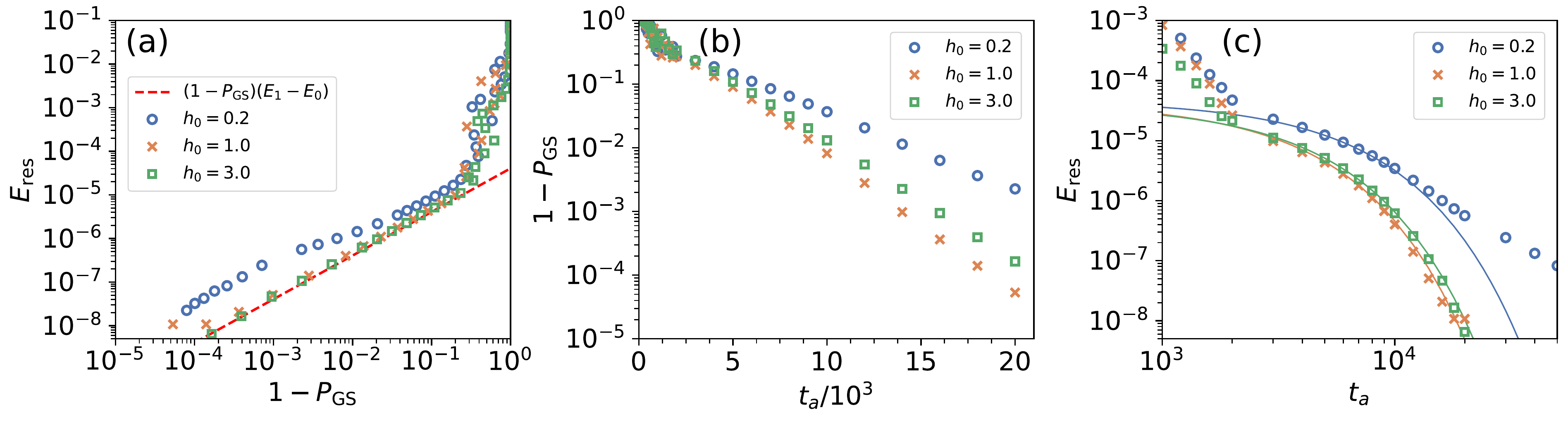}
  \caption{Results of TEBD simulations. (a) Relation between the residual energy and the deviation of the ground state probability from unity. The dashed line denotes $(1-P_{\mathrm{GS}})(E_1-E_0)$ where $E_1$ is the first excited energy of $H_0$. (b) Deviation of the ground state probability from unity as a function of the annealing time. (c) Residual energy and its fitting curve as a function of the annealing time. The solid lines represent the curves of $E_{\mathrm{res}}=C'\exp(-Ct_a)$ fitted to the data in the range of $3\times10^3 \leq t_a \leq 10^4$.}
  \label{fig:TEBD2}
\end{figure*}
\begin{figure}[t]
\centering
\includegraphics[width=0.73\hsize]{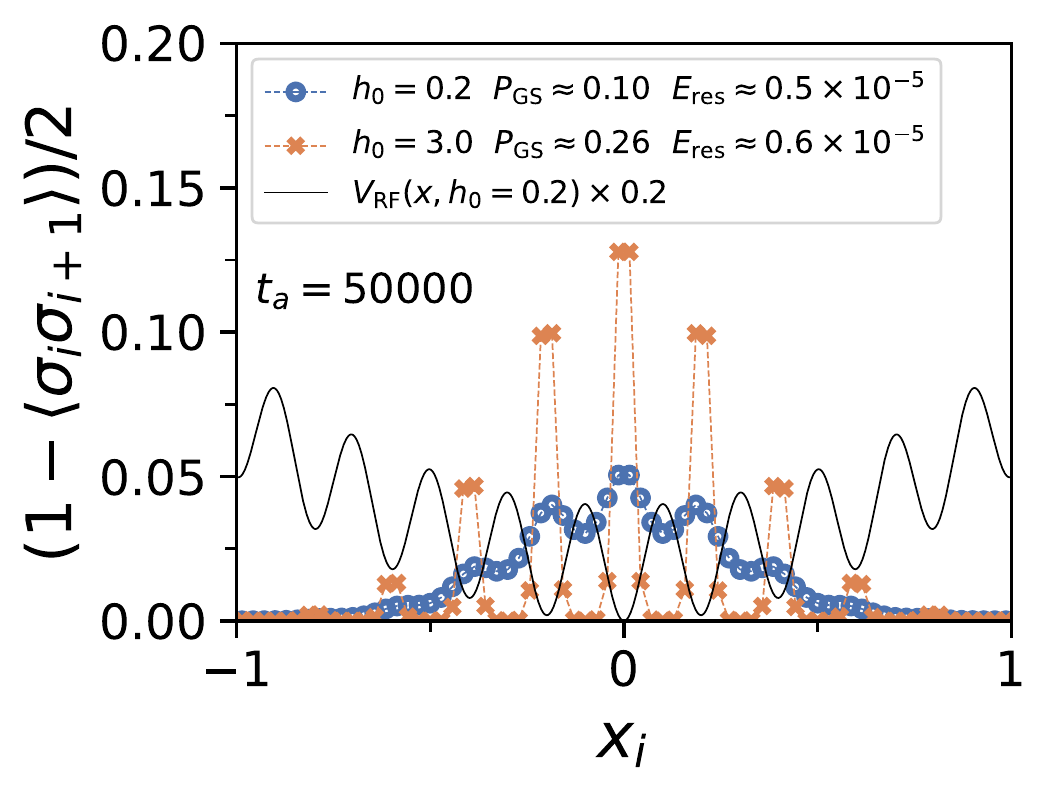}
\caption{Local number of kinks at $t_a=5\times10^4$ for $\lambda=0.01$ as a function of spatial coordinate $x_i=x_{\mathrm{min}}+i\Delta x $, where $i$ is the site index. The solid line represents the potential function of Eq.~\eqref{eq1} for $h_0=0.2$. The dashed lines are a guide to the eye.
}
  \label{fig:TEBD3}
\end{figure}
\subsection{\label{sec::sec42}TEBD for coherent QA}
In order to understand the origin of the suboptimal performance of the D-Wave described in the previous subsection, we now present the results of coherent QA based on TEBD to elucidate the differences between the D-Wave and the ideal QA that follows the Schr\"odinger dynamics without thermal noise.
The TEBD algorithm is based on the Suzuki-Trotter decomposition \cite{Suzuki_1979} of the time evolution operator and the matrix product state (MPS) representation to approximately follow the time evolution of the wave function \cite{Suzuki_2007, Ulrich_2011}. 
Time and energy are measured in the unit of $J$ in Eq. (\ref{eq5}) and we adopt the natural unit where $\hbar=1$. In this unit, if we choose $B(1)/2$ for the D-Wave as $J$, then $t_a=1$ corresponds to approximately 0.026 ns.
The time step size of the Suzuki-Trotter decomposition is set to $\Delta t=0.025/J$. The maximum bond dimension of the MPS, which controls the accuracy of the approximation, is $32$. We confirmed that the results converged well with respect to these parameters. 
 The annealing schedules are the linear function, as $A(s)/2=1-s$ and $B(s)/2=s$.

Figure \ref{fig:TEBD1} illustrates the results obtained by TEBD.
A significant difference from the experimental results by the D-Wave  in Fig.~\ref{fig:fig_4} is that the behavior of the kink density $\rho$ exhibits only very slight dependence on the height of the energy barrier $h_0$, which may originate in the weak dependence of coherent quantum tunneling effects on the barrier height. Moreover, in the case of coherent QA, for $t_a\geq10^3$, the single domain wall condition is almost completely satisfied, and the residual energy begins to decay quite rapidly, while in the case of the D-Wave in Fig.~\ref{fig:fig_4}, the single domain wall condition is not satisfied even in the long time regime, and the residual energy does not decay. These results suggest that deviations from the ideal Schr\"{o}dinger dynamics in the D-Wave lead to low performance, particularly for high $h_0$.
 
In Figs.~\ref{fig:TEBD1} (a) and (c), the kink density and the residual energy have the polynomial decay in the region $ t_a\leq 100$, and the fitting exponents are summarized in Table  \ref{table_1}. 
In the limit where $N$ is large and $\lambda$ is small, the theory of the Kibble-Zurek mechanism \cite{Kibble_1976, Zurek_1985, Dziarmaga_2005} predicts that $\rho\propto t_a^{-b}$ and $E_{\mathrm{res}}\propto t_a^{-c}$ with $b=c=1/2$ for the one-dimensional transverse-field Ising model without longitudinal fields. In the simulation for Fig.~\ref{fig:TEBD1}, $\lambda$ is set to 1, and therefore the magnitude of the longitudinal magnetic field is not negligible, see Eqs.~\eqref{eq8} and \eqref{eq11}, but the exponents show relatively mild dependence on $h_0$.

As shown in Figs.~\ref{fig:TEBD1} (b) and \ref{fig:TEBD1} (d), for $t_a>100$, the constraint satisfaction probability $P_{\mathrm{const}}$ starts to increase from 0 and the ground state probability $P_{\mathrm{GS}}$ also begins to increase. In this region, the residual energy deviates from the polynomial behavior and decreases more rapidly.

In a longer annealing time region ($t_a > 2\times10^3$), where the single domain wall condition is completely satisfied, $P_{\mathrm{GS}}$ approaches 1. In this regime, the energy expectation value can be expressed as 
\begin{align}
    \langle H_0\rangle = E_0 P_{\mathrm{GS}} + \mathcal{O}(1 - P_{\mathrm{GS}})\approx E_0 + \mathcal{O}(1 - P_{\mathrm{GS}}),
    \label{eq12}
\end{align}
and therefore the residual energy is proportional to $1-P_{\mathrm{GS}}$ as 
\begin{align}
E_{\mathrm{res}}=\frac{\langle H_0\rangle - E_0}{N} = \mathcal{O}(1-P_{\mathrm{GS}})\label{eq13}.
\end{align}
We can confirm this behavior by plotting the relation between $E_{\mathrm{res}}$ and $1-P_{\mathrm{GS}}$ in Fig.~\ref{fig:TEBD2} (a), as the slope is approximately 1 in the region where $1-P_{\mathrm{GS}}<0.1$.
The annealing time dependence of $1-P_{\mathrm{GS}}$ is shown in Fig.~\ref{fig:TEBD2} (b), and in the region where $1-P_{\mathrm{GS}}<0.1$, it can be seen that $1-P_{\mathrm{GS}}$ follows an exponential decay as $1-P_{\mathrm{GS}}\propto\exp(-Ct_a)$, which is reminiscent of the Landau-Zener formula. Consequently, the annealing time dependence of the residual energy can be expressed as $E_{\mathrm{res}} = C’\exp(-Ct_a)$, where $C'$ and $C$ are positive constants. The solid lines in Fig.~\ref{fig:TEBD2} (c) represent the result of fitting this expression to the data in the range of $3\times10^3 \leq t_a \leq 10^4$. In Ref.~\cite{Koh_2022}, an analysis of QA by direct numerical solution of the time-dependent Schr\"odinger equation for a single particle system in a continuous space showed that for a linear annealing schedule, the residual energy decays as the inverse square of the annealing time beyond the Landau-Zener regime as suggested in Ref.~\cite{Suzuki_2005}. The deviation of the residual energy for $h_0=0.2$ from the exponential behavior in the longer-time region $t_a>10^4$ may be an indication of a transition to such a power law behavior, although the exponent is much larger than two. To confirm the existence of this transition and to determine the power law exponents, simulations for much longer annealing times are necessary, which is challenging due to the accumulation of numerical errors in TEBD. 

\begin{figure*}[t]
\centering
\includegraphics[width=0.7\hsize]{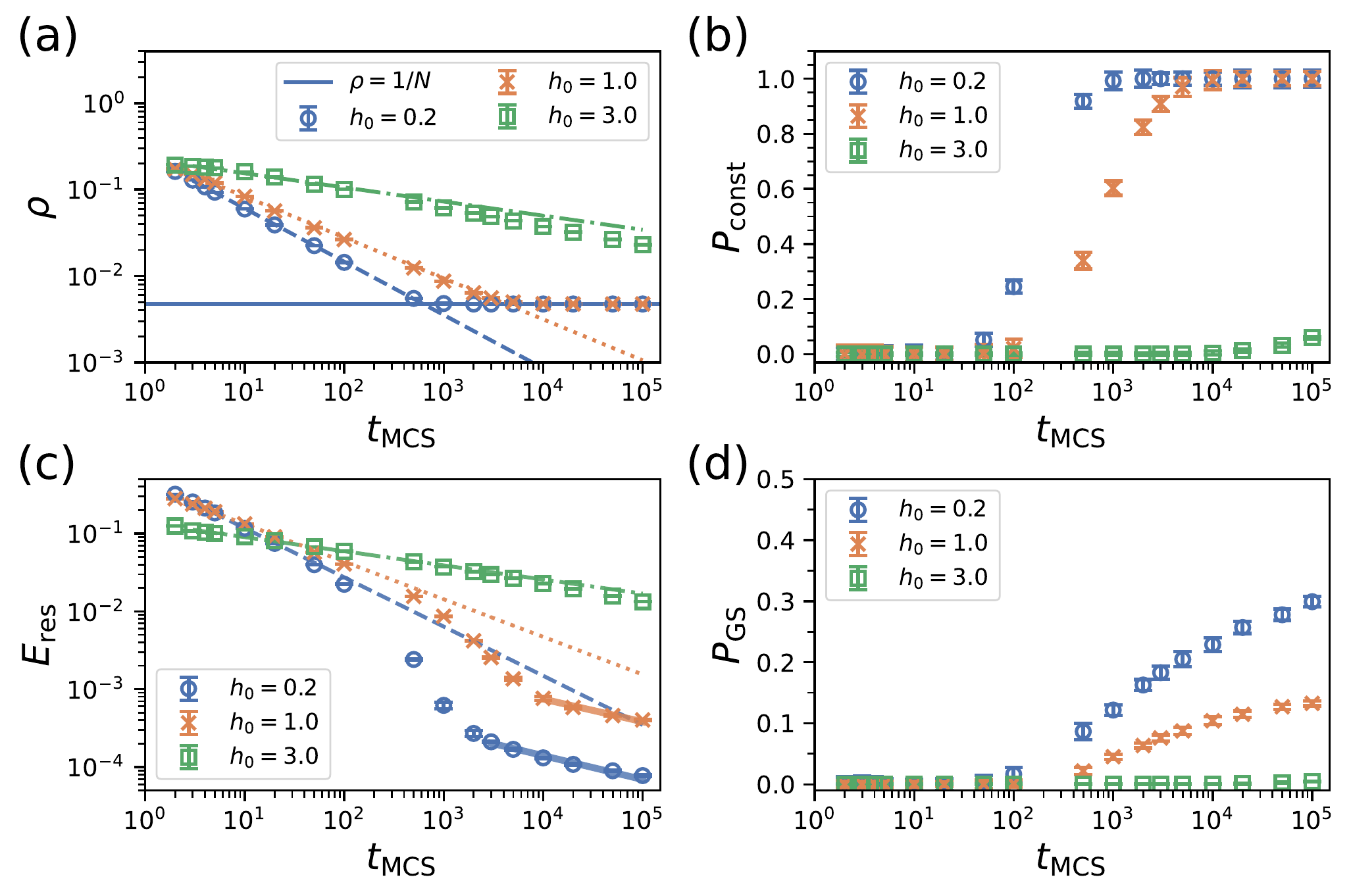}
  \caption{Dependence of the observables obtained from SA on the annealing time for different values of $h_0$.
  The vertical axes are the same as those in Fig. \ref{fig:fig_4}. 
  The horizontal axes represent the Monte Carlo step.
  }
  \label{fig:fig_5}
\end{figure*}
To understand the effect of the height of the energy barrier on the coherent QA, we analyze the spatial distribution of kinks in the wave function after the process of QA has terminated. Figure~\ref{fig:TEBD3} shows the local number of kinks $(1-\langle\hat{\sigma}_i^{z}\hat{\sigma}_{i+1}^{z}\rangle)/2$ at $t_a=5\times10^4$ as a function of spatial coordinate, as well as the values of the ground state probability $P_{\mathrm{GS}}$ and residual energy $E_{\mathrm{res}}$. For a small energy barrier $h_0=0.2$, kinks are distributed around the global minimum of the potential function at $x=0$ without significantly avoiding the maxima. This result suggests that for the low energy barrier, the kink is more likely to move around, making it difficult to remain at the bottom of the potential minima. On the other hand, for a large energy barrier $h_0=3.0$,  kinks localize in the vicinity of the minima and hardly exists in the vicinity of the maxima due to the high energy barrier.
The increase in the population at the local (but not global) minima for $h_0=3.0$ leads to larger residual energy than that of $h_0=0.2$. However, the probability of obtaining the ground state $P_{\mathrm{GS}}$ is higher when $h_0=3.0$. This behavior suggests that a lower energy barrier case is not necessarily easier to optimize, contrary to naive intuition. The characteristic of kink spatial distribution discussed here becomes indistinct when $\lambda$ is larger, but it can still be observed. Indeed, for $\lambda=1$, the optimization performance does not decrease even in the case of high energy barrier as shown in Fig.~\ref{fig:TEBD1}.

In summary, we have found that the optimization performance of coherent QA realized by TEBD is almost independent of the height of energy barrier $h_0$, and the system converges to the ground state  with high probability for sufficiently long annealing times. Therefore, the significant dependence of performance on $h_0$ and the premature saturation of the ground state probability in the data of the D-Wave can be attributed to imperfect realization of QA in the hardware, possibly mainly by thermal noise. As discussed in Appendix~\ref{appendix:thermal}, in order to fully understand the thermal effects on the hardware, it is necessary to analyze an open quantum system, not just an isolated system as done in this study. The performance of QA with a more coherent device would approach the results of TEBD simulations presented in this subsection and can be expected to be equivalent to or even better than that of global optimization algorithms such as BH and DE. Additional evidence will be provided in the following subsections to show that the $h_0$ dependence of data originates in thermal effects.

\subsection{\label{sec::sec43}Simulated annealing}
Next, we consider SA to investigate the effects of thermal fluctuations.
In SA, we use a single-spin update with the Metropolis rule. 
The temperature is linearly decreased as 
\begin{align}
T(t)=T_0+\frac{T_1-T_0 }{ t_{\mathrm{MCS}}-1 }\,t \quad(t=0,\dots,t_{\mathrm{MCS}}-1),
\end{align}
with $T_0=1$ and $T_1 =10^{-5}$.
The maximum number of Monte Carlo steps (MCS) is denoted as $t_{\mathrm{MCS}}$.
The number of spin updates is $t_{\mathrm{MCS}}N$. 
We repeat SA $20$ times  independently and sample $10^3$ spin configurations for each independent run.

\subsubsection{Kink density}
Figure~\ref{fig:fig_5} shows the results obtained by SA.
As in Fig.~\ref{fig:fig_4} (a), 
Fig.~\ref{fig:fig_5} (a) indicates the power law behaviors of the kink density in the short annealing time region $2\leq t_{\mathrm{MCS}}\leq100$.
Unlike the results obtained from the D-Wave  in Fig.~\ref{fig:fig_4} (a), the power law behaviors can be seen even in the case of high energy barrier, $h_0=3.0$.
For $h_0=0.2$, the exponents are larger than those of the results obtained by the D-Wave and the coherent QA, see Table \ref{table_1}.
On the other hand, for larger $h_0$ with higher energy barrier, the exponents are larger than those of the D-Wave and smaller than those of the coherent QA for the problem.
In this time region, the values of the kink density become larger than those of the D-Wave and take similar values to the case of coherent QA except for $h_0=3.0$.
For $h_0=3.0$, the kink density does not reach $\rho=1/N$ even at $t_{\mathrm{MCS}} = 10^5$.
 We need more annealing time to reduce the defects for the problem with the larger value of $h_0$.

\subsubsection{Single domain wall condition}
Figure \ref{fig:fig_5} (b) illustrates that few samples satisfy the single domain wall condition in the short-time region where the kink density has the polynomial decay. 
While this result is similar to the result obtained by the coherent QA in the short-time region as seen in Fig.~\ref{fig:TEBD1} (b), a clear  difference from  the D-Wave in Fig.~\ref{fig:fig_4} (b) is observed.
Since the system is not sufficiently equilibrated in the short annealing time region, many defects remain.
For the D-Wave result in Fig.~\ref{fig:fig_4} (b), $P_{\mathrm{const}}$ has a finite value except for $h_0=3.0$ in this region.
Although this D-Wave result is similar to the behavior of SA in the intermediate annealing time region $10^2\leq t_{\mathrm{MCS}}\leq10^3$ in Fig.~\ref{fig:fig_5} (b), it is impossible to reproduce all D-Wave data quantitatively by SA, for example, the residual energy as shown below.

In the long annealing time region  $t_{\mathrm{MCS}}\geq 10^4$, almost all samples satisfy the single domain wall condition except for $h_0=3.0$.
For $h_0=3.0$, $P_{\mathrm{const}}$ is suppressed. Similarly to the D-Wave data in Fig.~\ref{fig:fig_4} (b), SA is also affected by the height of the energy barrier.

\subsubsection{Residual energy}
As with the kink density in Fig.~\ref{fig:fig_5} (a), Fig.~\ref{fig:fig_5} (c) shows that the residual energy scales polynomially in the short annealing time region $2\leq t_{\mathrm{MCS}}\leq100$.
The exponents are smaller than the D-Wave results as seen in Table \ref{table_1} (b).
The exponents are larger than those by the coherent QA for $h_0=0.2$ and smaller 
 for $h_0=1.0$ and $3.0$.
These results imply that the dynamics based on thermal fluctuations are hampered by the higher energy barrier, and quantum fluctuations are more efficient than thermal fluctuations for higher barriers.

The residual energy of SA decreases polynomially in the long annealing time regions $2\times 10^3\leq t_{\mathrm{MCS}}\leq 10^5$ for $h_0=0.2$ and $10^4\leq t_{\mathrm{MCS}}\leq 10^5$ for $h_0=1.0$ where $P_{\mathrm{const}}\approx1$ holds. According to Table \ref{table_2}, the slope of the fitting line becomes smaller for the problem with the higher energy barrier.
The higher energy barrier impedes the transition to the ground states by thermal fluctuations.
On the other hand, the exponents obtained by the D-Wave show the opposite trend.
Note that $P_{\mathrm{const}}= 1$ does not hold for $h_0=1.0$ in Fig.~\ref{fig:fig_4} (b) for the D-Wave.
Whether or not this is related to the quantumness of the D-Wave is not clear at this point.
Although the exponents of SA are better than those of the D-Wave, the inverse square scaling of $E_{\mathrm{res}}$ is not observed.

\subsubsection{Ground state probability}
Figure \ref{fig:fig_5} (d) shows that the ground state probability starts to increase monotonically from $t_{\mathrm{MCS}}=10^2$. 
The performance depends on the height of the energy barrier.
The higher the energy barrier is, the smaller the success probability becomes.
The behavior is different from that in Figs.~\ref{fig:fig_4} (d) and \ref{fig:TEBD1} (d). 
Since the ground state probability is not close to 1 even in the long annealing time region, the exponential decay of $E_{\mathrm{res}}$ observed in the coherent QA in Fig.~\ref{fig:TEBD2} (c) does not appear in SA as seen in Fig.~\ref{fig:fig_5} (c).

\subsubsection{Summary}
The observables of SA clearly depend on $h_0$ while 
the observables of the coherent QA hardly depend on $h_0$. 
This is a significant difference between SA and coherent QA.
In coherent QA, quantum tunneling effects play an important role in the state transition between local minima. 
On the other hand, SA shows deteriorated performance under higher energy barriers. This behavior is consistent with the physical picture that SA goes over the energy barrier with thermal fluctuations.
Also in the case of the D-Wave, dependence of the observables on $h_0$ can be clearly seen.
Premature saturation of $P_{\mathrm{const}}$ and $P_{\mathrm{GS}}$ at lower values appear in the long annealing time region for the D-Wave.
Such saturation of these observables does not exist in SA.
This suggests that the degraded performance of the D-Wave may not be due to thermal noise alone, but to other factors as well \cite{Tristan_2021, Chancellor_2022}.

\begin{figure*}[t]
\centering
\includegraphics[width=0.7\hsize]{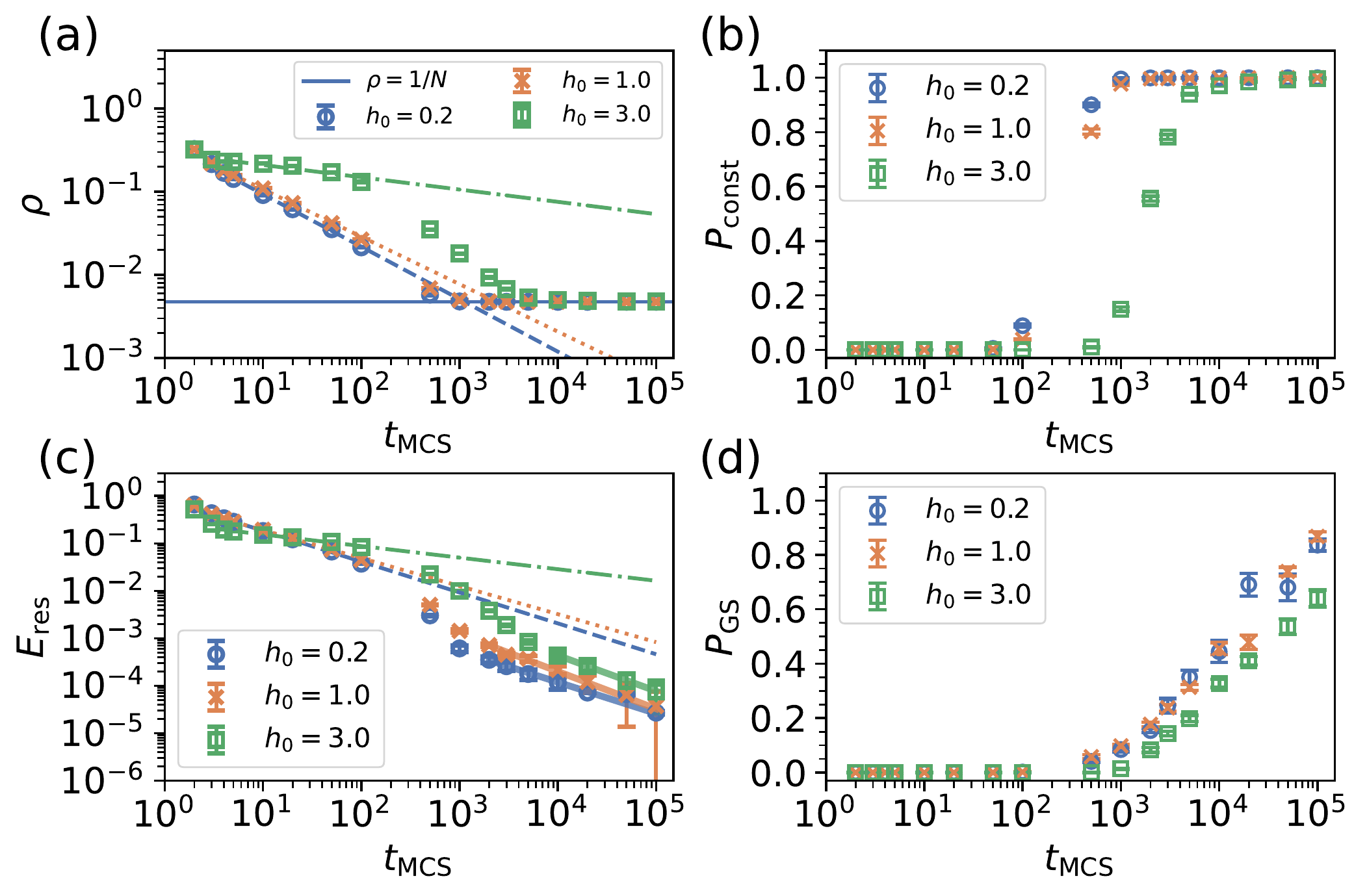}
  \caption{Dependence of the observables obtained by SQA on the annealing time for different values of $h_0$.
  Both axes are the same as those in Fig.~\ref{fig:fig_4}.
  }
  \label{fig:fig_6}
\end{figure*}
\subsection{\label{sec::sec44}Simulated quantum annealing}
Next, we study SQA and see its difference from SA and other protocols in some detail.
Simulated quantum annealing is the algorithm to simulate some aspects QA on a classical computer.
In SQA, we apply the Suzuki-Trotter decomposition \cite{Suzuki_1979} to the partition function of Eq.~\eqref{eq2}.
The quantum system at a finite temperature is then mapped to the effective classical Ising model as 
\begin{align}
&H_{\mathrm{eff}}= \nonumber\\
&\frac{B(s)}{2M}\sum_{k=1}^MH_0(k)+\frac{M}{2\beta}\ln \tanh \left(\frac{\beta A(s)}{2M}\right)\sum_{k=1}^M\sum_{i=1}^N\sigma_{ik}\sigma_{ik+1},
\label{eq16}\\
&H_0(k)=\lambda \sum_{i=1}^Nh_i\sigma_{ik}-J\sum_{i=1}^{N-1}\sigma_{ik}\sigma_{i+1k}+h\left(\sigma_{1k}-\sigma_{Nk}\right),
\label{eq17}
\end{align}
where $M$ is the Trotter number, $\beta=1/T$ is the inverse temperature, and $\sigma_{ik} \in\{\pm1\} (i=1,\dots,N,k=1,\dots,M)$ is the classical Ising configuration.
The coefficients of the  problem and transverse field are linearly controlled as 
\begin{align}
    \frac{A(s)}{2}&=1-s=1-\frac{t}{t_{\mathrm{MCS}}-1} ,\\
   \frac{B(s)}{2}&=s=\frac{t}{t_{\mathrm{MCS}}-1}, 
\end{align}
where $t=0,\dots,t_{\mathrm{MCS}}-1$.
We set the effective temperature $\beta/M=1$ and $M=10^3$. 
We utilize a single-spin update with the Metropolis rule.
We sweep the spin configurations $MN$ times for each MCS.
We perform SQA 20 times independently.
As shown in Ref.~\cite{Mario_2021}, the final spin configurations in Trotter replicas are utilized for computing the observables. 

\subsubsection{Kink density}
Figure~\ref{fig:fig_6} shows the results obtained from SQA.
In Fig.~\ref{fig:fig_6} (a), the kink density does not depend on the value of $h_0$ in the shortest annealing time region $2\leq t_{\mathrm{MCS}}\leq3$.
This behavior is similar to that of the coherent QA in Fig.~\ref{fig:TEBD1} (a).
In the short annealing time region $4\leq t_{\mathrm{MCS}}\leq100$, the polynomial decay of $\rho$ can be seen.
Except for $h_0=3.0$, SQA has the largest exponent in annealing protocols for discrete-value optimization, as shown in Table \ref{table_1}.
The exponents of $\rho$ become smaller as the height of the energy barrier increases.
The tendency is shared by other annealing protocols except for the coherent QA. 
For $h_0=1.0$ and $3.0$, SQA requires less annealing time to satisfy the single domain wall condition as compared to SA, and thus classically-simulated quantum fluctuations are more effective than thermal fluctuations in the higher barrier case.
This result reflects the difference in the dynamics of both methods and can also be understood from the behaviors of $P_{\mathrm{const}}$ in Fig.~\ref{fig:fig_6} (b).
For $h_0=0.2$, we cannot observe a clear difference between SA and SQA.
Compared with the coherent QA, the dependence of $\rho$ on $h_0$ appears.  The dynamics of SQA is fundamentally different from the dynamics of coherent QA \cite{Bando_2021} and may be regarded to be rather closer to the D-Wave behavior and SA.

\subsubsection{Single domain wall condition}
Figure~\ref{fig:fig_6} (b) shows that few samples satisfy the single domain wall condition in the short-time region similarly to the cases of SA and coherent QA in Figs.~\ref{fig:fig_5} (b) and  \ref{fig:TEBD1} (b).
The condition$P_{\mathrm{const}}\approx1$ holds in the long annealing time region $t_{\mathrm{MCS}}\geq10^4$ including the case of $h_0=3.0$. This behavior is different from the D-Wave and SA in Figs.~\ref{fig:fig_4} and \ref{fig:fig_5}.
No suppression of $P_{\mathrm{const}}$ for $h_0=3.0$ is observed for SQA, which is similar to coherent QA. 

\subsubsection{Residual energy}
Figure \ref{fig:fig_6} (c) illustrates that the residual energy decreases polynomially in the short annealing time region.
The range of fitting is the same as that in Fig.~\ref{fig:fig_6} (a) to extract the values in Table \ref{table_1}.
As seen in this Table, as the height of the energy barrier increases, the exponents of SQA decrease.
The exponents of SQA are larger than those of SA.
The result indicates that SQA can solve the problem more efficiently than SA.
For $h_0=3.0$, the exponent of SQA is smaller than that of the coherent QA. 
Although SQA can simulate some aspects of quantum effects, the higher energy barrier degrades the performance, unlike the coherent QA.
In the long annealing time region $10^3\leq t_{\mathrm{MCS}}\leq 10^4$, the polynomial decay of $E_{\mathrm{res}}$ can be observed.
As seen in Table \ref{table_2}, SQA can solve the problem more efficiently than the D-Wave and SA since SQA gives larger values of the exponent than the D-Wave and SA.
In our experiments, the exponential decay of $E_{\mathrm{res}}$ as in coherent quantum annealing is not observed in SQA. 
To see the behavior, we may need to investigate longer annealing time regions where $P_{\mathrm{GS}}\approx1$ holds.

\begin{figure*}[t]
\centering
\includegraphics[width=0.7\hsize]{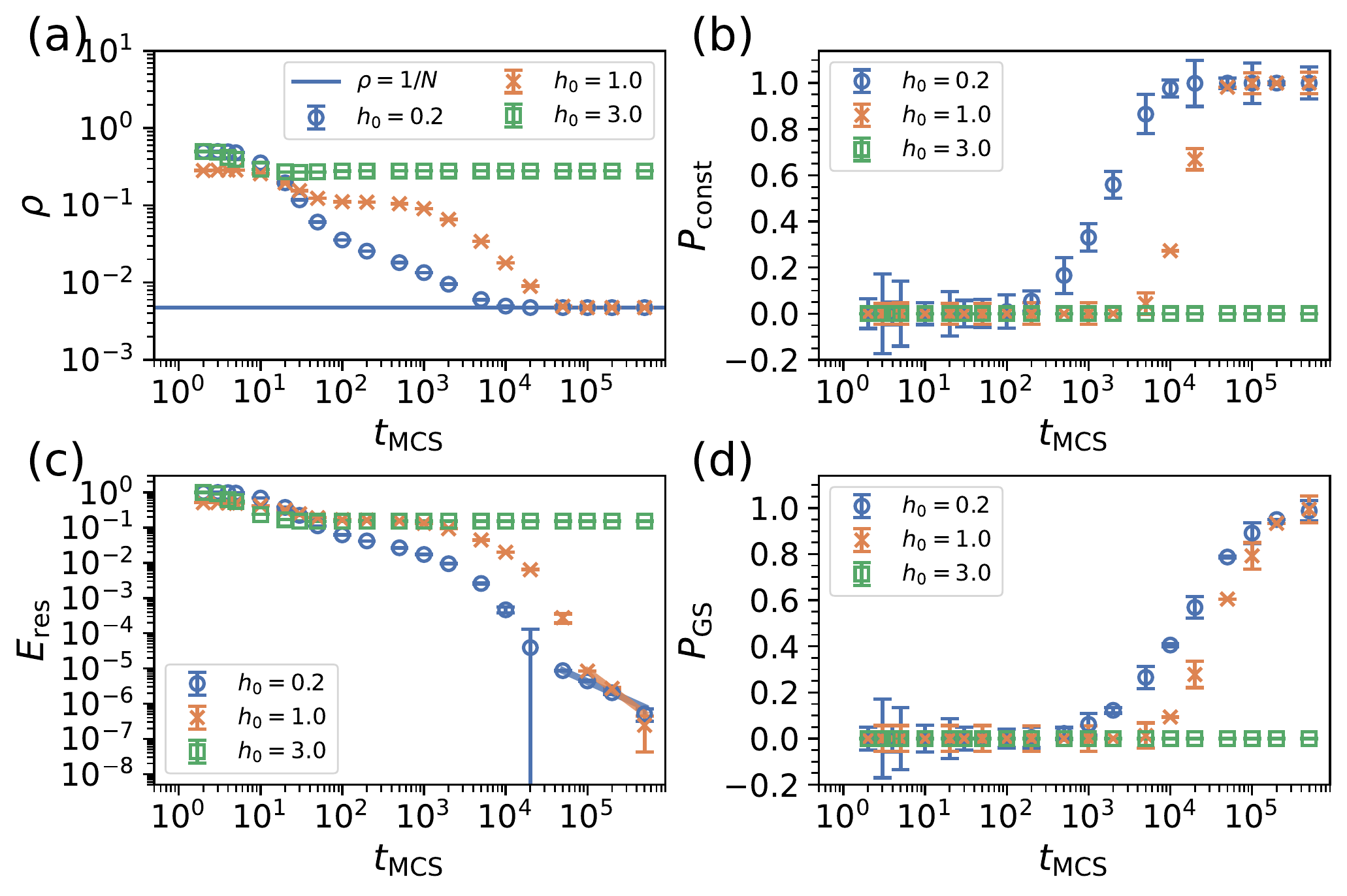}
  \caption{Dependence of the observables attained by SVMC on the annealing time for different values of $h_0$.
  Both axes are the same as those in Fig. \ref{fig:fig_4}.
  }
  \label{fig:fig_7}
\end{figure*}

\subsubsection{Ground state probability}
Figure \ref{fig:fig_6} (d) shows that the success probability becomes smaller as the height of the energy barrier increases in the large annealing time region $t_{\mathrm{MCS}}\geq 10^4$.
Since the values of $P_{\mathrm{GS}}$ obtained by SQA are larger than those by SA in the region $t_{\mathrm{MCS}}\geq 2\times10^3$, the effect of the energy barrier is weakened by classically-simulated quantum fluctuations.

\subsubsection{Summary}
The observables obtained by SQA depend on the value of $h_0$, which is different from the case of coherent QA.
This attributes to the fact that the dynamics of SQA  does not follow the Schr\"{o}dinger dynamics but is based on classical Monte Carlo updates under thermal fluctuations. 
Compared with SA, SQA suffers less performance degradation under the higher energy barrier, possibly by partly simulating tunneling effects \cite{Jiang2017}.

\subsection{\label{sec::sec45}Spin-vector Monte Carlo}
Spin-vector Monte Carlo is a classical algorithm proposed to explain the outputs from the D-Wave device \cite{Shin_2014}.
In SVMC, a quantum spin is  replaced by a classical rotor of unit length with a continuous angle variable $\theta_i\in [0,\pi]$ as
\begin{align}
    \hat{\sigma}_i^z\rightarrow \cos \theta_i, \quad\hat{\sigma}_i^x\rightarrow \sin \theta_i.
\end{align}
The total Hamiltonian can then be described as :
\begin{align}
H&=\frac{B(s)}{2}H_0-\frac{A(s)}{2}\sum_{i=1}^N\sin \theta_i,\label{eq18}\\
H_0&=\lambda \sum_{i=1}^Nh_i\cos \theta_i-J\sum_{i=1}^{N-1}\cos \theta_i \cos \theta_{i+1}\nonumber\\
&+h\left(\cos \theta_1 -\cos \theta_N \right)\label{eq19}.
\end{align}
The annealing schedules of $A(s)/2$ and $B(s)/2$ are the same as those in SQA. 
The temperature in Monte Carlo process is set in our simulation to $T=10^{-5}$ and the number of all spin updates is written as $Nt_{\mathrm{MCS}}$.
Each continuous angle is initialized to $\theta_i=\pi/2$, and spins are randomly chosen and updated with the Metropolis rule as $\theta_i\rightarrow\theta_i'\in [0,\pi]$. 
In addition, we apply the transverse-field-dependent (TFD) update \cite{Albash_2021, Bando_2022}, which is proposed to capture the freezing phenomena in the D-Wave device.
In the TFD update, the current angle is updated as $\theta_i\rightarrow \theta_i +\epsilon_i(s)$  
where $\epsilon_i(s)$ is the random variable dependent on the annealing parameter and the energy coefficients,
\begin{align}
   \epsilon_i(s) \in \left[-\mathrm{min}\left(1,\frac{A(s)}{B(s)}\right)\pi,\mathrm{min}\left(1,\frac{A(s)}{B(s)}\right)\pi\right].  
\end{align}
The range of updating the angle gradually decreases in the late annealing process $A(s)<B(s)$.
We execute SVMC $10$ times independently and sample $10^3$ spin configurations for each run.
\subsubsection{Kink density}
Figure~\ref{fig:fig_7} shows the results of SVMC.
In Fig.~\ref{fig:fig_7} (a), nonlinear behaviors of $\rho$ are observed.
In the region of very short annealing time $2\leq t_{\mathrm{MCS}}\leq 3$, the kink density does not depend on the values of $h_0$.
Dependence of $\rho$ on the values of $h_0$ appears for $t_{\mathrm{MCS}}\geq 3$.
For $h_0=3.0$, the kink density saturates at a larger value than that of SA in the region $t_{\mathrm{MCS}}\geq 50$.
This result represents that SVMC is more affected by the high energy barrier than SA.
\subsubsection{Single domain wall condition}
Figure~\ref{fig:fig_7} (b) shows that most samples satisfy the single domain wall condition $t_{\mathrm{MCS}}\geq 10^5$ for $h_0=0.2$ and $1.0$.
Compared with other annealing protocols,
SVMC needs more annealing time to satisfy $P_{\mathrm{const}}\approx1$.
The result indicates that the dynamics of SVMC is slower than that of other annealing protocols.
\subsubsection{Residual energy}
As in $\rho$ in Fig.~\ref{fig:fig_7} (a), Fig.~\ref{fig:fig_7} (c) illustrates that nonlinear behavior of $E_{\mathrm{res}}$ emerges and the saturation of $E_{\mathrm{res}}$ for $h_0=3.0$ appears. 
In the long annealing time region, the polynomial decay of $E_{\mathrm{res}}$ can be observed. 
The fitting regions are $5\times10^4\leq t_{\mathrm{MCS}}\leq 5\times10^5$ for $h_0=0.2$ and $10^5\leq t_{\mathrm{MCS}}\leq 5\times10^5$ for $h_0=1.0$.
In these regions, $P_{\mathrm{const}}\approx1$ holds as seen in Fig.~\ref{fig:fig_7} (b).
From Table \ref{table_2}, the exponents of SVMC are the largest among all annealing protocols. Therefore, SVMC can reach the ground states faster than other annealing protocols in the long time region.
Compared with SQA in the region where the single domain wall condition holds, the values of $E_{\mathrm{res}}$ obtained by SVMC are smaller for $h_0=0.2$ and $1.0$. 
\subsubsection{Ground state probability}
Figure~\ref{fig:fig_7} (d) demonstrates that the ground state probability starts to increase more slowly than SA and SQA.
The higher energy barrier suppresses the increase of $P_{\mathrm{GS}}$.
At a later stage, SVMC produces a larger value of $P_{\mathrm{GS}}$ than SQA does for $h_0=0.2$ and $1.0$ in the region where $P_{\mathrm{const}}\approx1$ holds.
\begin{figure*}[t]
\centering
\includegraphics[width=0.7\hsize]{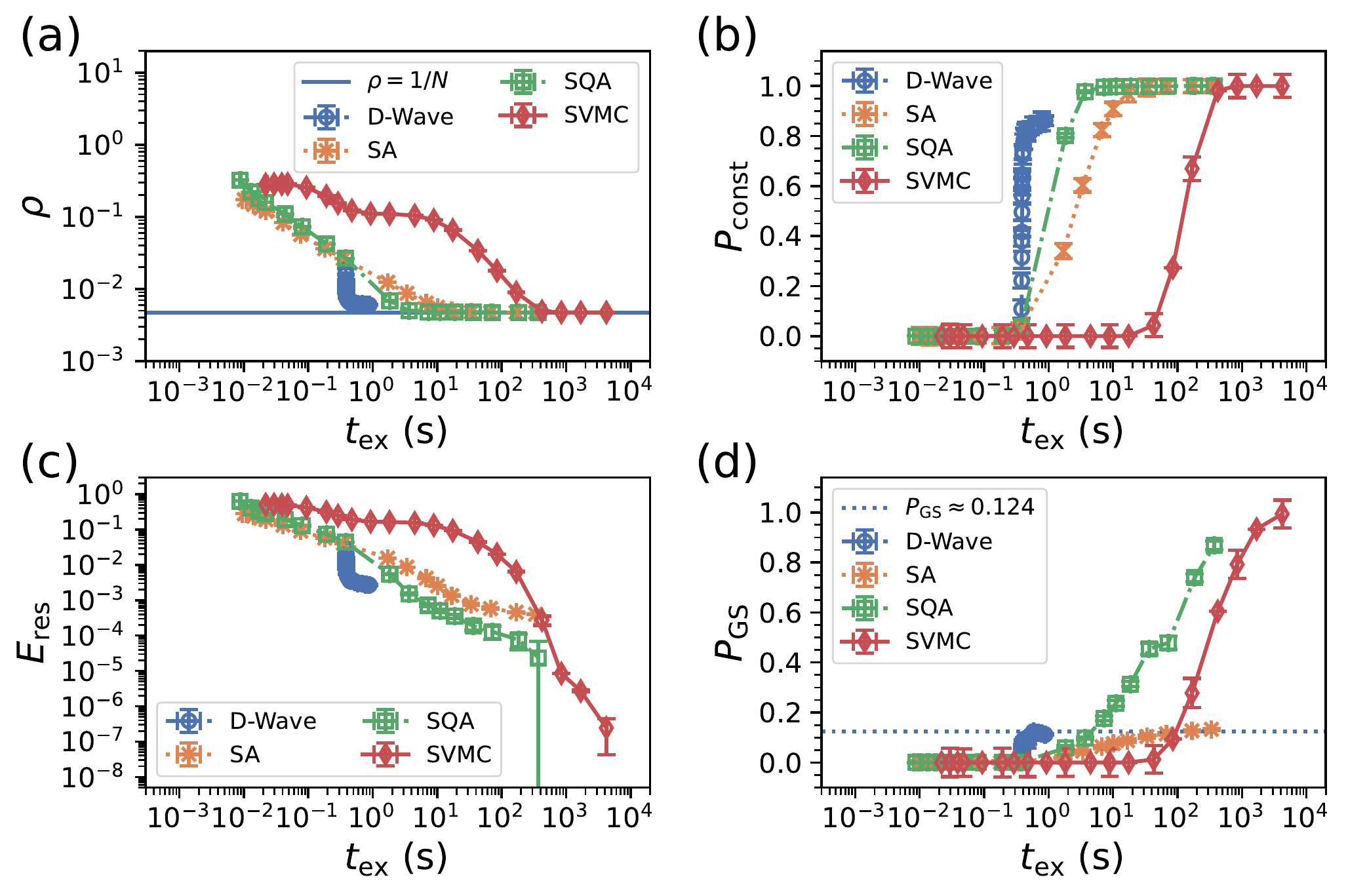}
  \caption{Dependence of the observables on the execution time  for  different annealing protocols with $k=0.5$, $w_0=0.2$, and $h_0=1.0$.
Both axes are the same as those in Fig.~\ref{fig:fig_4}. 
  }
  \label{fig:fig_11}
\end{figure*}
\subsubsection{Summary}
Values of observables in SVMC depend on the value of $h_0$ as in SA and SQA.
The dynamics of SVMC is more affected by the higher energy barrier. 
This is a consequence of the fact that the dynamics of SVMC is activated by thermal fluctuations in the semi-classical potential function.
Although the behaviors of the observables are unique to SVMC and are different from those obtained by the D-Wave, the performance of SVMC is the best in the region where $P_{\mathrm{const}}\approx1$ holds except for the coherent QA. 
This result suggests that the dynamics of SVMC is more effective than other classical annealing protocols for the Rastrigin function in the space where only a single domain wall exists. 

Considering SVMC as a practical solver for continuous function optimization, a smaller absolute value of the execution time is required, not just larger scaling exponents. In the next section, we analyze the dependence of observables on the absolute execution time. 
\subsection{\label{sec::sec46}Actual execution time}
Finally, we show the dependence of observables on the actual execution time in seconds for different annealing protocols in Fig.~\ref{fig:fig_11}.
We use the same data as those presented in this Sec.~\ref{sec::sec4} for $k=0.5$, $w_0=0.2$, and $h_0=1.0$.
The definition of $t_{\mathrm{ex}}$ is the same as that defined in Sec. \ref{sec::sec32}.
Since the TEBD is the deterministic algorithm and different from the other annealing protocols based on the Monte Carlo method, we avoid to show the TEBD data for direct comparison in this section.

Figure~\ref{fig:fig_11} (a) illustrates that SQA reaches $\rho=1/N$ most quickly among different annealing protocols
\footnote{
Notice that the D-Wave becomes stuck at a value larger than $\rho=1/N$ as shown in Fig.~\ref{fig:fig_4} (a) though this behavior is not clearly seen in the scale of Fig.~\ref{fig:fig_11} (a).}. 
Although the D-Wave reaches the lowest value of $\rho$ around  $0.4 ({\rm sec})\leq t_{\mathrm{ex}}\leq 0.9 ({\rm sec})$, its value of $\rho$ remains suboptimal in the longer time region due to the noise and other imperfections as observed more clearly in Fig.~\ref{fig:fig_4} (a).  

Around the moderately-short time region $0.4 ({\rm sec})\leq t_{\mathrm{ex}}\leq 0.9 ({\rm sec})$, $P_{\mathrm{const}}$ reaches close to 1 by the D-Wave  as shown in Fig.~\ref{fig:fig_11} (b).
If the goal is to satisfy the condition of single domain wall as quickly as possible in the actual time of execution, the D-Wave is the choice.

Figure~\ref{fig:fig_11} (c) shows that the D-Wave  has the smallest value of $E_{\mathrm{res}}$ around $0.4 ({\rm sec})\leq t_{\mathrm{ex}}\leq 0.9 ({\rm sec})$.
Around $2 ({\rm sec})\leq t_{\mathrm{ex}}\leq 370 ({\rm sec})$, SQA performs best. 
From the analysis of the exponents in Table \ref{table_2},  SVMC surpasses SQA in the long time region, and the difference of SQA and SVMC widens with time.
To satisfy the single domain wall condition, SVMC needs more time, see Figs.~\ref{fig:fig_11} (a) and (b).  However, once the condition is satisfied, the residual energy decreases most rapidly by SVMC.

Figure~\ref{fig:fig_11} (d) indicates that the success probability of the D-Wave  has the largest value around $0.4 ({\rm sec}) \leq t_{\mathrm{ex}}\leq0.9 ({\rm sec})$,  
$P_{\mathrm{GS}}\approx 0.124$ as indicated by the horizontal line. 
The D-Wave  reaches this value most quickly among all annealing protocols. However, it shows saturation at around this finite value, and the performance hardly improves with time.
In Fig.~\ref{fig:fig_2}, we see that the ground state probability obtained by BH and DE are large $P_{\mathrm{GS}}>0.9$ even in the longest time region $10 ({\rm sec})\leq t_{\mathrm{ex}}\leq100 ({\rm sec})$.
As far as our implementation is concerned, other annealing protocols, SA, SQA, SVMC,  are similarly less effective than the global optimization algorithms for continuous-variable optimization since the former do not achieve such high values in this long-time region.

Overall, in the short time window of $0.4 ({\rm sec}) \leq t_{\mathrm{ex}}\leq0.9 ({\rm sec})$, the D-Wave  is the best choice.  At around the intermediate time $t_{\mathrm{ex}} \approx 2 ({\rm sec})$, SQA is competitive.  Beyond this time region, dedicated global optimization algorithms for continuous variables,  DE and BH, keep improving with time, whereas classical metaheuristics for discrete optimization remain relatively mediocre. The D-Wave stays stagnant.  It is nevertheless expected from the data of TEBD that the ideal QA without noise is likely to be most competitive, if realized at the hardware level, also in the intermediate and long-time regions.

\section{\label{sec:sec5}Conclusion}

We have performed extensive benchmarks of QA and classical optimization algorithms for the continuous-variable Rastrigin function in one dimension with a rugged energy landscape.
By using the domain wall encoding, the continuous variable problem was represented by a one-dimensional Ising chain.

We have first compared the D-Wave with classical algorithms for continuous-variable optimization with local or global updates and have demonstrated quantitatively that, in the case of higher energy barrier, the performance of the D-Wave is better or comparable to that of the local optimization algorithms in a limited region of execution time and is comparable to the global optimization algorithms in the same time region. For longer execution times, the global algorithms show outstanding performance whereas the D-Wave remains stagnated.
For the potential function with lower energy barrier, no clear advantage of the D-Wave has been observed. While the performance of the global optimization algorithms keeps improving with increasing execution time, no improvements were identified in the D-Wave.

For better understanding of the performance of QA, in its hardware realization of the D-Wave as well as in an ideal setting without noise, we next compared QA with protocols for discrete-variable optimization, TEBD for coherent QA, SA, SQA, and SVMC.
The observed values of physical quantities show a clear dependence on the height of the energy barrier except for TEBD for coherent QA.  This is natural for SA, SQA, and SVMC because these classical algorithms make explicit use of thermal noise to go over the barrier. In the case of the D-Wave, its barrier-height dependence implies that the process realized in the D-Wave is explicitly affected by thermal noise. Particularly in the case of the highest barrier with $h_0=3.0$, we found no sign of the successful solution by the D-Wave, whereas coherent QA by TEBD reached the ground state with high probability for relatively long annealing time with little-to-no dependence on the barrier height. This result suggests that the performance degradation of the D-Wave at long annealing time  may be significantly reduced by improvements in the hardware to reduce noise and possibly other imperfections as demonstrated in recent experiments of quantum simulation \cite{King_2022, King2022b}. Stated otherwise, quantum annealing has the potential to become quite competitive even for continuous-variable problems under the domain wall encoding. This is achieved by quantum tunneling effects, as implied by the independence of the barrier height observed in the TEBD data. Classical algorithms, including those for continuous-variable optimization, would struggle to find a solution if the barrier become even higher than we tested, while QA, if realized coherently, would be much less affected by the height. This conclusion has become possible by the analysis of the continuous-variable problem with an explicit and easy control of the barrier height, which is generally difficult to achieve in problems expressed directly by the Ising model.  In other words, we have demonstrated the effectiveness of coherent quantum tunneling effects explicitly and quantitatively in an optimization problem. Simulated quantum annealing may be regarded as a quantum-inspired classical algorithm that partly takes into account quantum effects. Nevertheless, comparison of data for TEBD in Fig.~\ref{fig:TEBD1} and SQA in Fig.~\ref{fig:fig_6} suggests that a complete realization of quantum effects in QA is necessary for better results.

This paper represents a first step toward systematic and quantitative studies of continuous-variable optimization by QA in comparison with a series of well-established classical algorithms.  In particular, the present one-dimensional problem with a periodic structure of the barrier is one of the simplest examples of continuous-variable optimization.  The global optimization algorithm DE based on the genetic algorithm may find it more difficult to reach the correct solution for aperiodic potentials than for the periodic Rastrigin function. Also, higher-dimensional problems with the complex energy landscape as studied qualitatively in Refs.~\cite{Abel_2021a,Abel_2021b,Abel_2022}, which are generally difficult to analyze by TEBD, may be the interesting next target of systematic studies.

\section*{Acknowledgment}
The work of H.N. is based on a project JPNP16007 commissioned by the New Energy and Industrial Technology Development Organization (NEDO).  

\appendix
\section{Classical algorithms for continuous-variable optimization \label{appendix_a}}
In this Appendix, we explain in some detail the classical optimization algorithms used for benchmarking.
We consider the optimization problem as 
\begin{align}
    \min_{\bm{x}\in\mathbb{R}^n} f(\bm{x}),
\end{align}
where the objective function is defined by $f:\mathbb{R}^n\rightarrow \mathbb{R}$.
The implementation of all classical algorithms for benchmarking utilizes  Scipy \cite{Scipy2020}.
\subsection{Nelder-Mead}
The Nelder-Mead (NM) method is commonly used for unconstrained continuous optimization problems \cite{Nelder_1965,Gao_2012}.
The NM method does not need any information on derivatives and utilizes only the observed functional value. 
In the NM method, we hold a simplex constructed from the convex hull of $n+1$ points in $\mathbb{R}^n$.
We iteratively update the simplex by removing the worst functional point and substituting another better point for it.
The candidate is selected by the following process: reflecting, expanding, or contracting the simplex along the line connecting the worst point with the centroid of the other points \cite{NoceWrig06}.
If the solution cannot be improved anymore, the simplex is shrunk as all other points get closer to the current best solution.
The detailed algorithm of the NM method is explained in Refs.~\cite{NoceWrig06,Gao_2012}.

\subsection{Conjugate gradient descent}
The conjugate gradient decent (CGD) method is an iterative algorithm with the gradient of the objective function \cite{NoceWrig06}.
The original algorithm is designed for solving a linear system of equations \cite{Hestenes_1952}. The extended algorithm for nonlinear functions was proposed by Fletcher and Reeves \cite{Fletcher_1964}. In the CGD method, we set the initial point $\bm{x}^{(1)}\in\mathbb{R}^n$ and its direction $\bm{d}^{(1)}=-\nabla f(\bm{x}^{(1)})$, and update the current point $\bm{x}^{(t)}$ as follows: $\bm{x}^{(t+1)}\leftarrow\bm{x}^{(t)}+\alpha^{(t)}\bm{d}^{(t)}$ where $\alpha^{(t)}$ is the step length obtained via a line search for minimizing $f(\bm{x}^{(t)})$ along the direction $\bm{d}^{(t)}$.
The current direction is calculated by using the gradient as $\bm{d}^{(t+1)}\leftarrow -\nabla f(\bm{x}^{(t+1)})+\beta^{(t+1)}\bm{d}^{(t)}$. 
In the Fletcher and Reeves formulations,
the hyperparameter $\beta^{(t+1)}$ is determined by $\beta^{(t+1)}\leftarrow\nabla f(\bm{x}^{(t+1)})^T\nabla f(\bm{x}^{(t+1)})/\nabla f(\bm{x}^{(t)})^T\nabla f(\bm{x}^{(t)})$. We iterate the above procedure until the algorithm converges to a local minimum or the number of iterations reaches the maximum iteration.
\subsection{Basin-Hopping}
The Basin-hopping (BH) method is a widely used global optimization algorithm for nonlinear multimodal optimization problems, where the local optimization algorithm often fails to find the global minimum due to the rugged energy landscape \cite{Leary_2000}.
The BH method is combined with the gradient-based local optimization and the perturbation based on Monte Carlo random walks with the Metropolis rule \cite{Iwata_2004,Luo_2012}.
Starting from an initial point $\bm{x}^{(1)}$, we apply the local optimization to map a current point $\bm{x}^{(t)}$ to the nearest minima $\bm{y}^{(t)}$ in the local optimization process. 
Next, we add a perturbation to the current solution as $\bm{x}_{\mathrm{new}}=\bm{y}^{(t)}+\bm{\epsilon}$ where $\bm{\epsilon}$ is a random coordinate generated from the uniform distribution $U(a,b)$. 
We perform the local optimization once again and attain another local minimum $\bm{y}_{\mathrm{new}}$.
Following the Metropolis rule, we accept the proposed point as $\bm{y}^{(t+1)}=\bm{y}_{\mathrm{new}}$ or reject it as $\bm{y}^{(t+1)}=\bm{y}^{(t)}$ 
with the probability, $\mathrm{min}\left[1,\exp(-(f(\bm{y}_{\mathrm{new}})-f(\bm{y}^{(t)}))/T)\right]$ where $T$ is the temperature to adjust the scale of the energy. 
The above iteration continues until $t=t_{\mathrm{max}}$. 
In our experiments, we utilize the default setting $a=-0.5$, $b=0.5$, and $T=1$, and adopt the CGD method in the local optimization process.

\subsection{Differential evolution}
The differential evolution (DE) method is a stochastic search algorithm for minimizing real-valued global optimization problems based on the evolutionary strategy \cite{Storn_1996, Storn_1997}.
In DE, we initialize a population vector for each generation $G$ as $\bm{x}_i^G\in \mathbb{R}^n \hspace{2pt} (i=1,\dots,p, G=0,\dots,G_{\mathrm{max}})$, where $p$ is the number of populations.
The population vector is generated from the uniform distribution to cover the entire parameter space. 
Next, we create a new population vector with different ones as $\tilde{\bm{x}}_i^{G+1}={\bm{x}}_{r_1}^G+F\left({\bm{x}}_{r_2}^G-{\bm{x}}_{r_3}^G\right)$, where $r_1,r_2,$ and $r_3$ are indices of the population determined by the \textit{``crossover"} rule and  $F>0$ is the amplification factor.   
The new trial vector $\bm{u}_i^{G+1}$ can be obtained by mixing the target vector $\bm{x}_i$ and the new population vector $\tilde{\bm{x}}_i^{G+1}$ with the \textit{``crossover"} rule \cite{Das_2010}. 
If the new trial vector has a lower objective function value than the target vector, the new trial vector supersedes the target vector and is inherited by the next generation.
The above process is called the \textit{``selection"}.
The algorithm stops at $G=G_{\mathrm{max}}$ or the mean objective function value over all populations becomes lower than the threshold $\epsilon$.
In our experiments, we adopt the \textit{``best/2/bin"} strategy \cite{Storn_1996, Storn_1997} and set $\epsilon=10^{-7}$.
The number of populations $G_{\mathrm{max}}$ corresponds to the maximum number of iterations $t_{\mathrm{max}}$ in the main text.
The other hyperparameters are used in the default settings.

\section{Additional data for QA}
\label{appendix_b}
In this Appendix, we show some additional data from the D-Wave and consider the problem with $k=0.5$, $w_0=0.2$, and $h_0=1.0$. 
We compute the observables as illustrated in Sec.~\ref{sec::sec32}.
The experimental setting of the D-Wave is almost the same as those in Fig.~\ref{fig:fig_4}.
 \begin{figure*}[t]
\centering
\includegraphics[width=0.7\hsize]{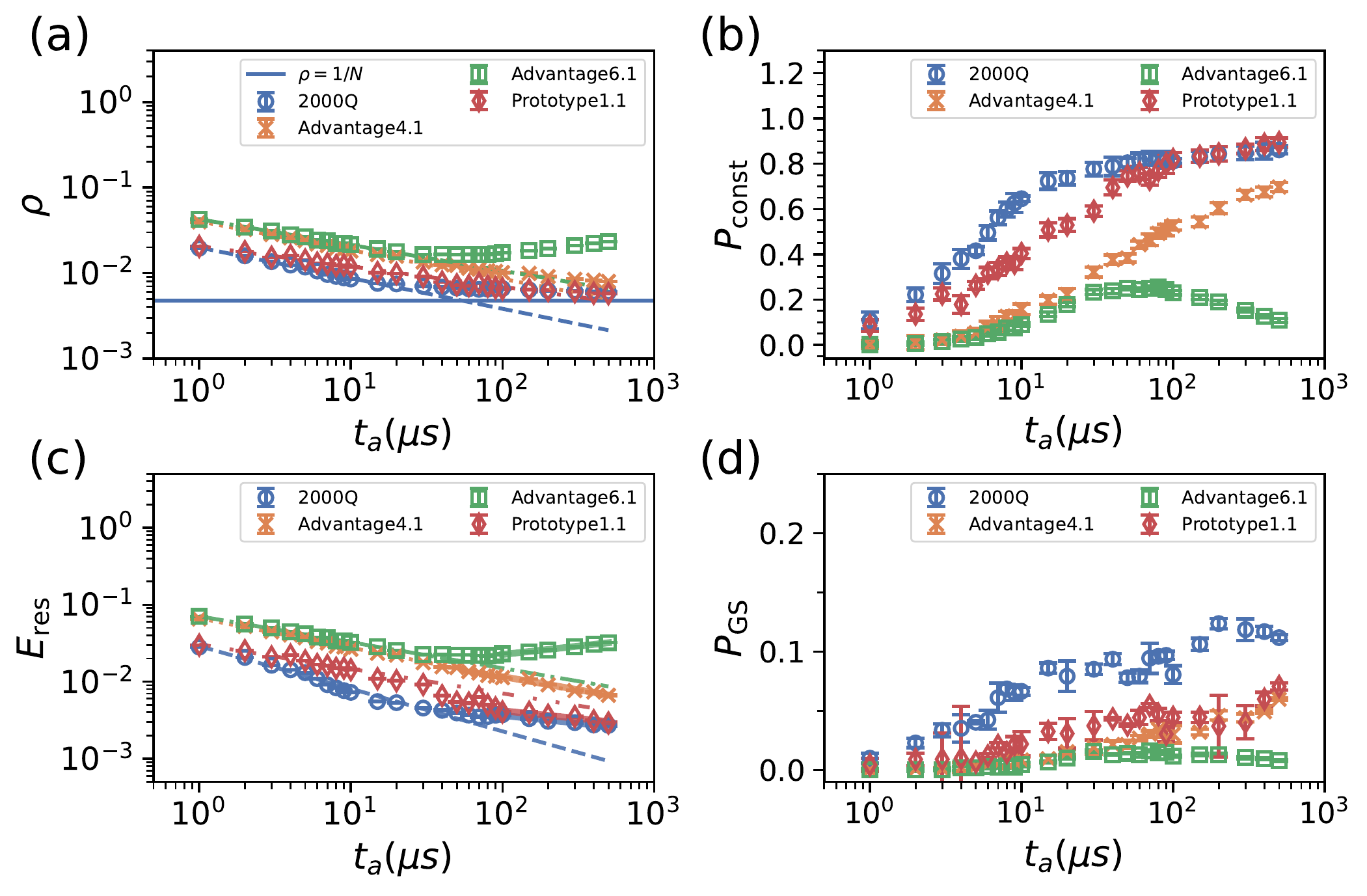}
  \caption{Dependence of the observables on the annealing time for the different quantum annealers.
  Both axes are the same as those in Fig. \ref{fig:fig_4}. 
  }
  \label{fig:fig_13}
\end{figure*}
\begin{table*}[htbp]
  \centering
  \caption{The fitting exponents of the observables for the different quantum annealers. 
   The "$E_{\mathrm{res}}$ (short)" represents the residual energy  in the short annealing time region $1\leq t_a\leq 10$.
The "$E_{\mathrm{res}}$ (long)" is the residual energy in the long annealing time region $100\leq t_a\leq 500$.}
  \begin{tabular}{|c||c|c|c|c|}  \hline
    machine &2000Q&Advantage4.1&Advantage6.1&Prototype1.1  \\ \hline 
    $\rho$ &  
    $0.359\pm0.011$&
    $0.337\pm0.008$& 
    $0.301\pm0.005$&
    $0.239\pm0.015$\\ \hline
    $E_{\mathrm{res}}$ (short) &  
    $0.553\pm0.025$&
    $0.383\pm0.012$&
    $0.338\pm0.006$&
    $0.319\pm0.021$\\ \hline
    $E_{\mathrm{res}}$ (long)&  
    $0.201\pm0.02$&
    $0.341\pm0.036$&
    $-0.218\pm0.012$&
    $0.201\pm0.031$\\ \hline
  \end{tabular}
  \label{table_4}
\end{table*}
\subsection{Comparison of D-Wave 2000Q and D-Wave Advantage}
\label{appendix_diff_annealer}
Figure~\ref{fig:fig_13} illustrates the dependence of observables on the annealing time for a few different quantum annealers.
In addition to the ``DW\_2000Q\_6" device, we utilize the ``Advantage\_system 4.1", ``Advantage\_system 6.1", and ``Advantage2\_prototype 1.1" devices and compare their performance. 

Regardless of the type of annealers, the polynomial decay of $\rho$ can be observed in the short annealing time region $1(\mu s)\leq t_a\leq10(\mu s)$ as shown in Fig.~\ref{fig:fig_13} (a). 
The fitting exponents are summarized in Table \ref{table_4}.
The largest exponent of $\rho$ is obtained from the D-Wave 2000Q. 
Except for the D-Wave Advantage 6.1, the kink density gradually decreases as we increase the annealing time.
The kink density obtained by the D-Wave Advantage 6.1 shows similar behavior to the D-Wave Advantage 4.1 in the region $1(\mu s)\leq t_a\leq10(\mu s)$ and starts to increase at $t_a=30$ $(\mu s)$. 
This behavior may be due to decoherence.
In the long annealing time region $t_a\geq 100(\mu s)$, no device reaches the target $\rho=1/N$.
In this region, the D-Wave 2000Q takes similar values as the D-Wave Advantage2 Prototype 1.1 does.

Figure~\ref{fig:fig_13} (b) illustrates that  the D-Wave 2000Q has the largest $P_{\mathrm{const}}$ in the region $1(\mu s)\leq t_a\leq 100(\mu s)$.
The D-Wave Advantage2 Prototype 1.1 takes similar values of $P_{\mathrm{const}}$ to those obtained by the D-Wave 2000Q in the region $100(\mu s)\leq t_a\leq500(\mu s)$.
In the long annealing time region, most samples obtained by the above two devices satisfy the single domain wall condition.
Both devices yield the larger value of $P_{\mathrm{const}}$ than the D-Wave Advantage 4.1 and 6.1.

Figure~\ref{fig:fig_13} (c) shows that the residual energy decays polynomially irrespective of the types of the quantum annealer 
in the short annealing time region as also seen in Fig.~\ref{fig:fig_13} (a).
From Table~\ref{table_4}, the largest exponent of $E_{\mathrm{res}}$ can be obtained by the D-Wave 2000Q. 
In our experiments, the D-Wave 2000Q gives the lowest value of $E_{\mathrm{res}}$ among all devices.
In the long annealing time region $t_a\geq100(\mu s)$, we can see a polynomial decay of $E_{\mathrm{res}}$.
The fitting exponents obtained by the D-Wave 2000Q and the D-Wave Advantage2 prototype 1.1 take similar values to those in Table \ref{table_4}.
The result indicates that qualitative behaviors of the two devices 
show little differences.
The D-Wave Advantage 4.1 has the largest exponent among all devices, from which one may expect Advantage 4.1 to yield smaller values of $E_{\mathrm{res}}$ than other models if the device would operate with minimal noise in the longer annealing time region.

Figure~\ref{fig:fig_13} (d) demonstrates that the D-Wave 2000Q gives the largest value of $P_{\mathrm{GS}}$.
Although the D-Wave 2000Q and the D-Wave Advantage2 Prototype 1.1 give similar exponents in $E_{\mathrm{res}}$ in the long annealing time region from Table \ref{table_4}, the performance gap appears in $P_{\mathrm{GS}}$.
This result implies that the D-Wave Advantage2 Prototype 1.1 may reach low energy states, but not necessarily the ground states. 

The observable values obtained by different quantum annealers show different behaviors.  We generally conclude that the D-Wave 2000Q is suitable for optimizing the Rastrigin function under the present problem setting.

\begin{figure*}[t]
\centering
\includegraphics[width=0.7\hsize]{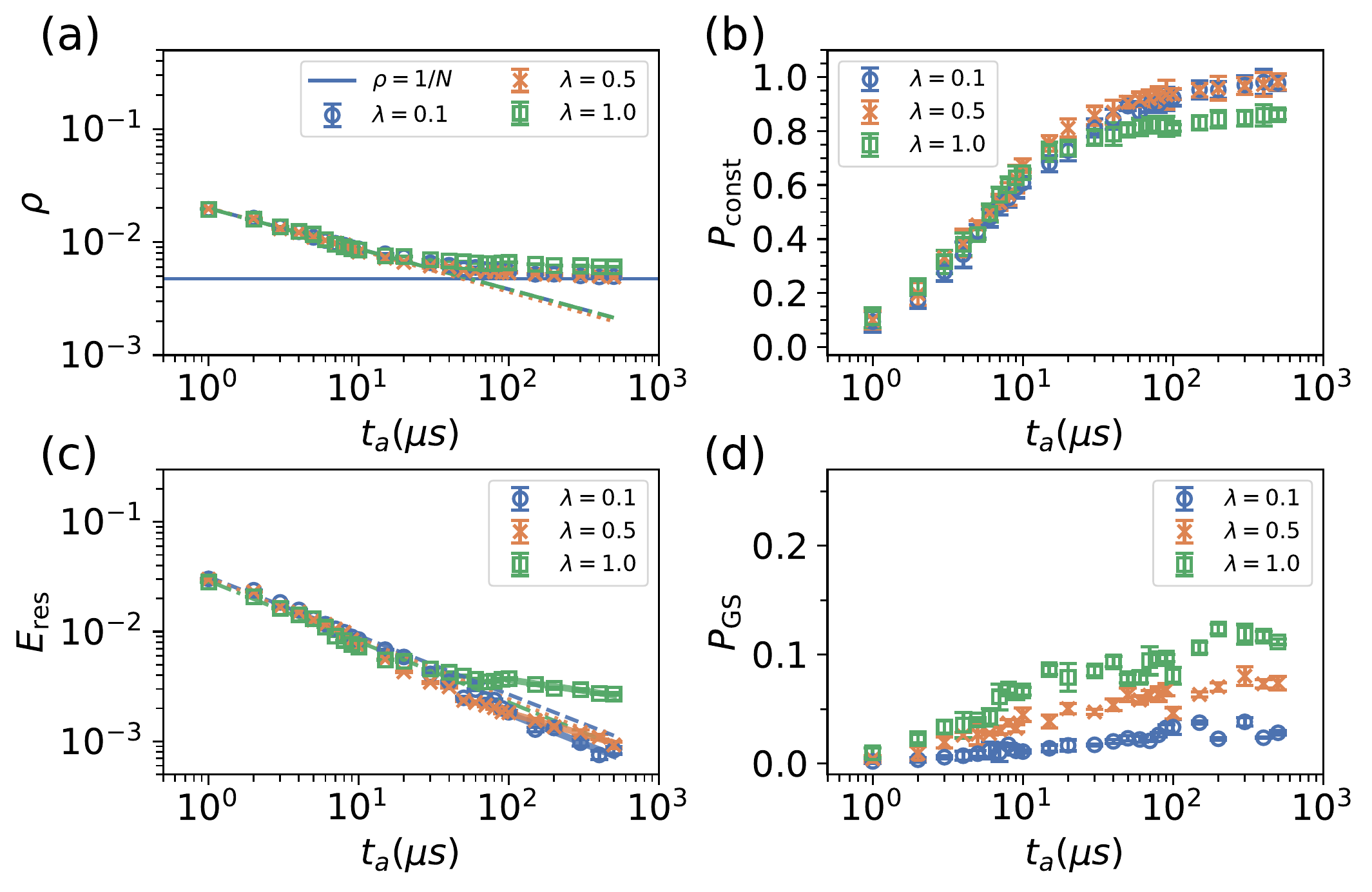}
  \caption{Dependence of the observables attained by  the D-Wave on the annealing time for different values of $\lambda$.
  Both axes are the same as those in Fig. \ref{fig:fig_4}. 
  }
  \label{fig:fig_12}
\end{figure*}
\label{appendix_b1}
\begin{table*}[htbp]
  \centering
  \caption{The fitting exponents of the observables for different values of $\lambda$.
 The fitting region is the same as used in Table  \ref{table_4}.}

  \begin{tabular}{|c||c|c|c|}  \hline
    $\lambda$ & $0.1$&$0.5$&$1.0$  \\ \hline 
    $\rho$ &  $0.359\pm0.011$&$0.372\pm0.011$&  $0.359\pm0.011$\\ \hline
    $E_{\mathrm{res}}$ (short) &  $0.536\pm0.026$&$0.553\pm0.024$&  $0.553\pm0.025$\\ \hline
      $E_{\mathrm{res}}$ (long) &  $0.545\pm0.073$&$0.392\pm0.019$&  $0.201\pm0.02$\\ \hline
  \end{tabular}
  \label{table_3}
\end{table*}
\subsection{Dependence of the observables for different values of $\lambda$}
\label{appendix_b2}
Figure~\ref{fig:fig_12} shows the dependence of the observables on the annealing time for different values of $\lambda$.
All data are obtained from the D-Wave 2000Q. 
As in the main text, 
we refer to the D-Wave 2000Q as the D-Wave.
The data for $\lambda=1.0$ is the same as used in  Fig.~\ref{fig:fig_4}.
In Fig.~\ref{fig:fig_12} (a), the polynomial decay of $\rho$ can be seen in the short annealing time region $1(\mu s)\leq t_a\leq10(\mu s)$ for different values of $\lambda$. 
The fitting exponents are summarized in Table \ref{table_3}. 
In this region, the coefficient $\lambda$ hardly affects the behavior of the system. 
On the other hand, in the long annealing time region $t_a\geq 100(\mu s)$, the saturation seen in $\rho$ for $\lambda=1.0$ does not occur in $\lambda=0.1$ and $0.5$. 
The small $\lambda$ weakens the effect of the problem term Eq.~\eqref{eq9} and  enhances the effect of the constraint term  Eq.~\eqref{eq5}.

Figure~\ref{fig:fig_12} (b) demonstrates 
the value of $\lambda$ hardly affects $P_{\mathrm{const}}$ in the short annealing time region.
Similarly to the kink density in Fig.~\ref{fig:fig_12} (a), the effect of $\lambda$ can be seen in the long annealing time region $t_a\geq 100(\mu s)$.
Premature finite saturation of $P_{\mathrm{const}}$ appears for $\lambda=1.0$.
The number of samples that satisfy the single domain wall condition increases by adjusting $\lambda$.
In the long annealing time region, the effect of quantum fluctuation becomes weaker due to the decoherence. In this region,  thermal fluctuations affect the system.
Since small values of $\lambda$ largely reduce the height of the energy barrier and the lower energy barrier is favorable for thermal fluctuations as shown in Fig.~\ref{fig:fig_5}, the number of samples that satisfy the single domain wall condition increases with decreasing $\lambda$ in this region.

Figure~\ref{fig:fig_12} (c) shows that the polynomial decay of $E_{\mathrm{res}}$ can be observed in the short annealing time region $1(\mu s)\leq t_a\leq10(\mu s)$.
From Table~\ref{table_3}, 
the values of $\lambda$ hardly affects the exponents of $E_{\mathrm{res}}$ as in the kink density. In the long annealing time region $t_a\geq 100(\mu s)$, the residual energy decreases polynomially and the fitting exponents are written in Table~\ref{table_3} (b). 
A smaller $\lambda$ yields a larger exponent. 
This result implies that a smaller $\lambda$ leads to lower energy states and reduces the hardness of optimization due to the higher energy barrier.

Figure~\ref{fig:fig_12} (d) illustrates that larger values of $\lambda$ give larger $P_{\mathrm{GS}}$. 
This result is in contrast to the behavior of $E_{\mathrm{res}}$ in Fig.~\ref{fig:fig_12} (c). 
Smaller values of $\lambda$ decrease the values of the magnetic field physically encoded into the D-Wave.
Since the bit width to represent the problem Hamiltonian is limited in the D-Wave, the actual implemented value in the D-Wave may be different. 
Due to this discretization error, the actual energy landscape slightly changes.
Because the success probability is more affected by variation of the energy landscape than the residual energy, the success probability for the smaller $\lambda$ is degraded.

\section{Comparison of the data of the D-Wave 2000Q and thermal equilibrium state \label{appendix:thermal}}
The D-Wave is a device to reproduce the system behavior under the time dependent Hamiltonian $H(s)$ in Eq.~(\ref{eq2}). However, as the hardware is an open system, its dynamics is affected by thermal noise. To describe such dynamics, the hypothesis of quasi-static time evolution and freeze-out phenomenon has been proposed \cite{Amin_2015}. In this hypothesis, during the annealing process, the density matrix of the system at a given time is approximately equal to the density matrix of the thermal equilibrium state of the Hamiltonian at that time. The dynamics freezes at a time denoted as $s^*$, when the relaxation time to equilibrium becomes much larger than the time scale of annealing, and the density matrix of the final result becomes $\exp(-H(s^*)/T_{\mathrm{phys}})$, where $T_{\mathrm{phys}}$ denotes the physical device temperature that is about 13.5 mK. If the strength of the transverse field at the freeze-out point, $A(s^*)$, is much smaller than the classical part of the Hamiltonian, $B(s^*)$, then the density matrix of the system can be approximated as 
\begin{align}
    \rho_{\mathrm{sys}}(s^*) \approx
 \frac{e^{-(B(s^*)/2)H_0/T_{\mathrm{phys}}}}{\mathrm{Tr}[e^{-(B(s^*)/2)H_0/T_{\mathrm{phys}}}]}. 
\end{align}
This can be interpreted as the density matrix at the thermal equilibrium state of $H_0$ with a temperature of $2T_{\mathrm{phys}}/B(s^*)$. 
By determining this rescaled temperature from the output samples of the D-Wave, it is possible to estimate the freeze-out point $s^*$. It should be noted that if the obtained effective temperature value is too high, the estimation of $s^*$ may become inaccurate and it would be appropriate to estimate the temperature using $H(s^*)$ instead of $H_0$.

\begin{figure}[t]
\centering
\includegraphics[width=0.8\hsize]{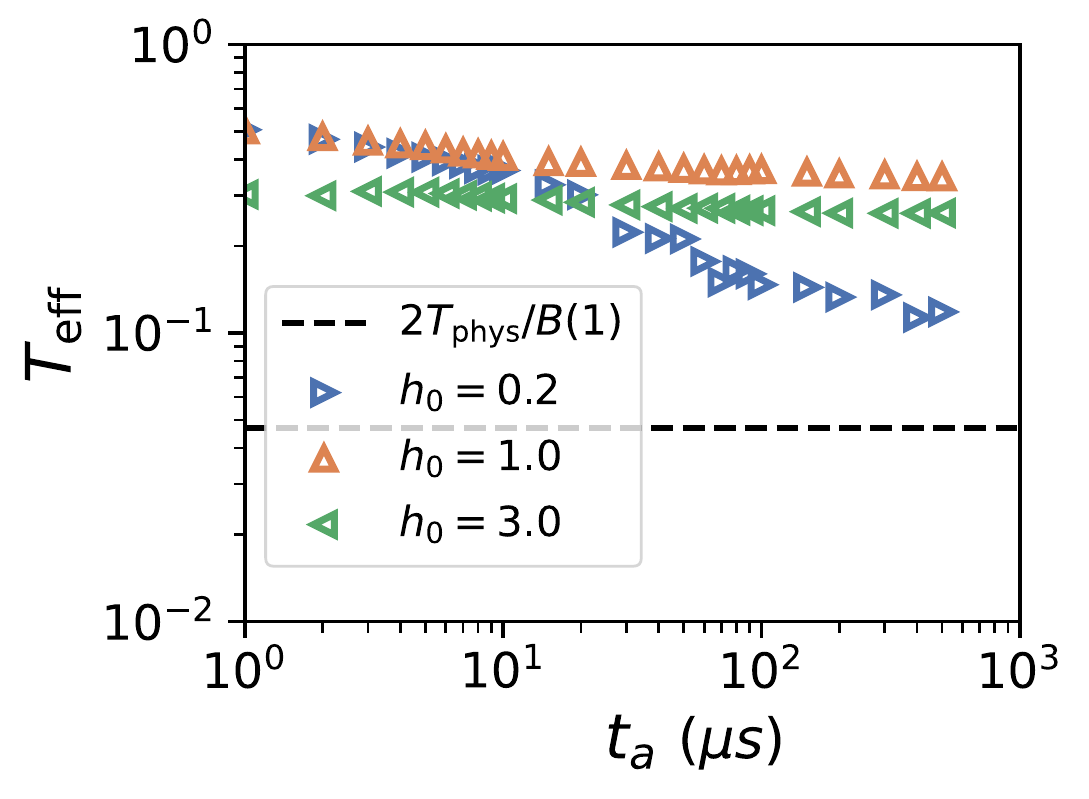}
  \caption{Effective temperature obtained from the data of  the D-Wave as a function of annealing time. The dashed horizontal line represents the physical temperature of the D-Wave.}
  \label{fig:app_c_Teff}
\end{figure}

We now discuss the consistency between the experimental results of  the D-Wave in this study and the hypothesis of quasi-static time evolution and freeze-out. The effective temperature $T_{\mathrm{eff}}$ is defined as the value obtained by solving the equation $E(T) = \langle H_0 \rangle$ for $T$, where $E(T)$ is the internal energy of the thermal equilibrium state of $H_0$ at temperature $T$ and $\langle H_0 \rangle$ is the energy expectation value of the data of  the D-Wave. The annealing time dependence of the effective temperature from the experimental results of  the D-Wave is shown in Fig.~\ref{fig:app_c_Teff}. For $h_0=0.2$, the effective temperature decreases toward $T_{\mathrm{phys}}$ while for $h_0=1.0$ and $3.0$, the effective temperature saturates at a constant value several times higher than $2T_{\mathrm{phys}}/B(1)$.

\begin{figure}[t]
\centering
\includegraphics[width=0.8\hsize]{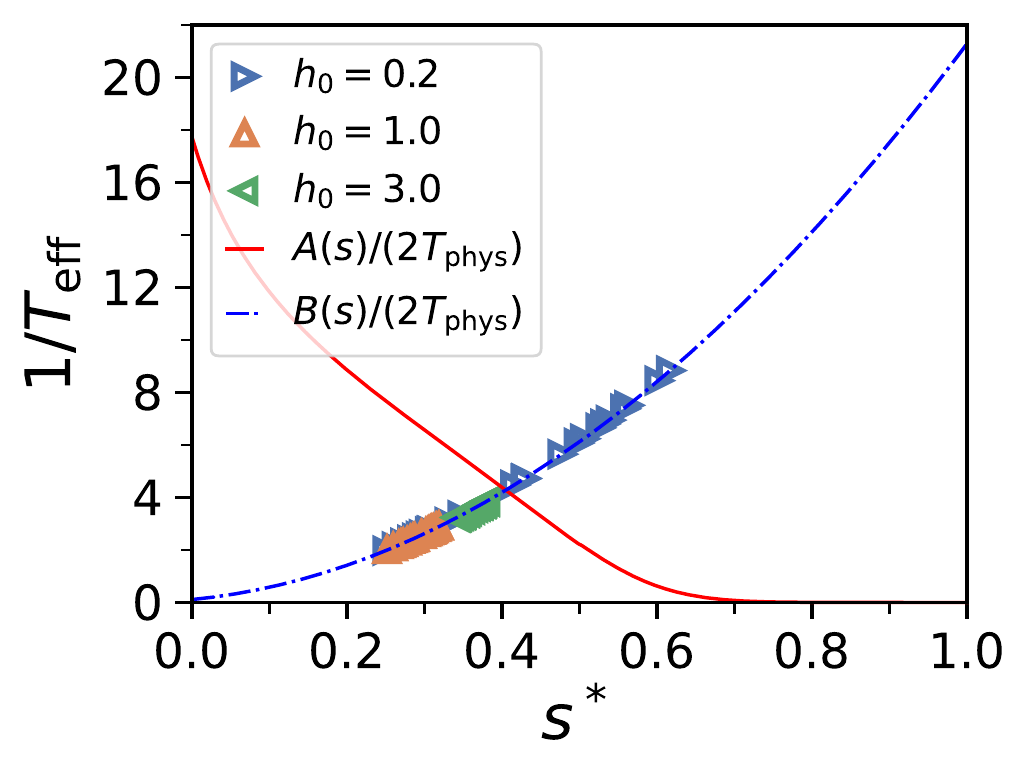}
  \caption{Relation between the inverse effective temperature and freeze-out point. The lines represent the annealing schedule functions used to determine $s^*$ from $T_\mathrm{eff}$.}
  \label{fig:app_c_schedule}
\end{figure}
Regarding $T_{\mathrm{eff}}$ as $2T_{\mathrm{phys}}/B(s^*)$, the freeze-out point $s^*$ is estimated as the value obtained by solving the equation $2T_{\mathrm{phys}}/B(s)=T_{\mathrm{eff}}$ for $s$. The relation between $s^*$ and $1/T_{\mathrm{eff}}$ is shown in Fig.~\ref{fig:app_c_schedule} and the annealing schedule function $B(s)$ used to determine $s^*$ and the schedule function of the transverse field $A(s)$ are also shown. For $h_0=0.2$, at $t_a=500 (\mu s)$ the freeze-out point $s^*$ reaches approximately 0.6, where $B(s^*)\gg A(s^*)$, indicating that the classical term is dominant in $H(s^*)$. 
For $h_0=1.0$ and $h_0=3.0$, $s^*$ reaches approximately 0.35 and 0.4, respectively. For these $s^*$ values, $B(s^*)\approx A(s^*)$ and $B(s^*)< A(s^*)$, respectively, which may lead to the deviation of the data of the D-Wave from the thermal equilibrium of $H_0$.

\begin{figure*}[t]
\centering
\includegraphics[width=0.8\hsize]{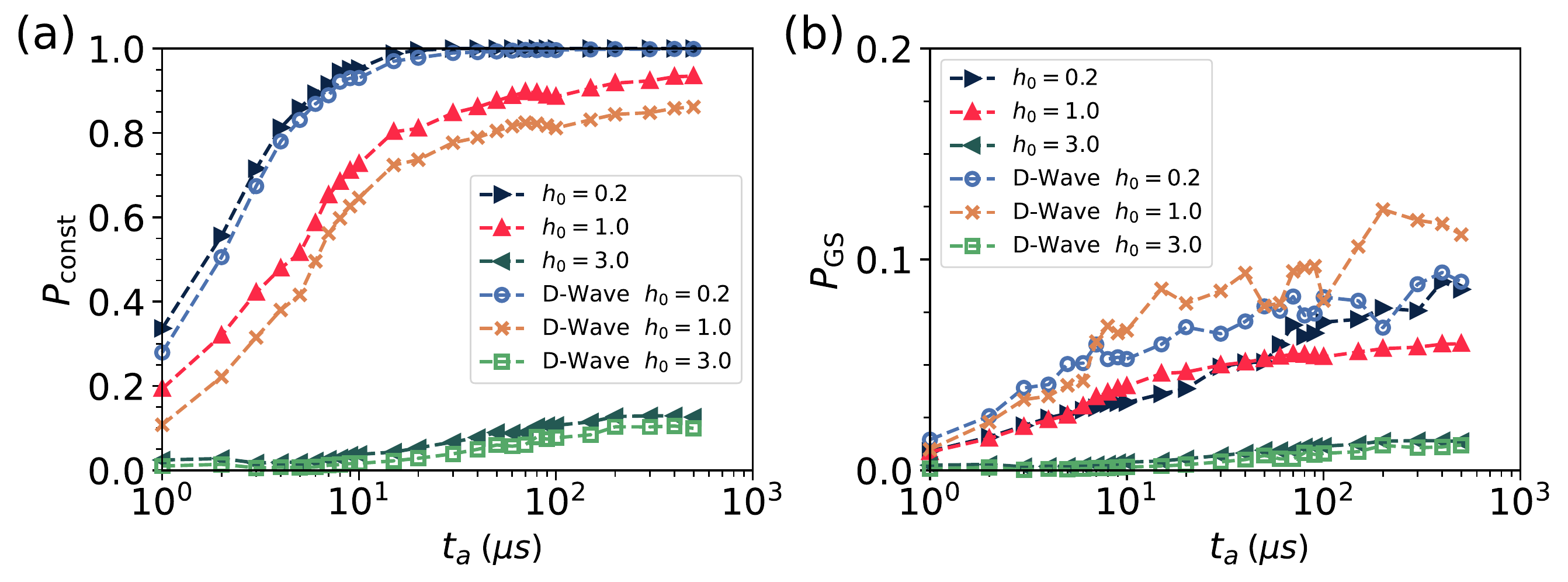}
  \caption{Comparison between data of  the D-Wave (shown by symbols with lines) and the thermal equilibrium of $H_0$ at the corresponding effective temperature (shown by filled triangles). (a) Constraint satisfaction probability and (b) ground state probability as functions of annealing time.}
  \label{fig:app_c_DW}
\end{figure*}

\begin{figure*}[t]
\centering
\includegraphics[width=0.8\hsize]{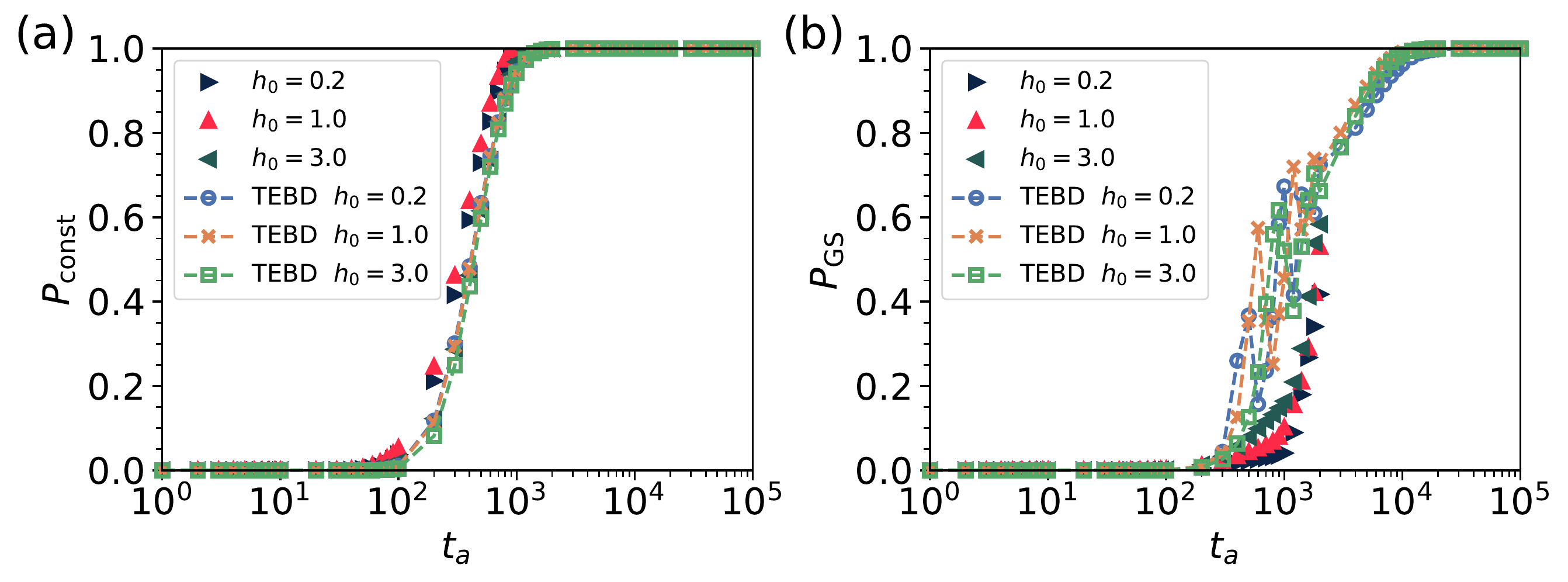}
  \caption{Comparison between data of coherent QA obtained by TEBD (shown by symbols with lines) and the thermal equilibrium of $H_0$ at the corresponding effective temperature (shown by filled triangles). (a) Constraint satisfaction probability and (b) ground state probability as functions of annealing time. The annealing time $t_a$ is in the unit of $J$.}
  \label{fig:app_c_TEBD}
\end{figure*}

To examine how closely the D-Wave data matches a thermal equilibrium value, we compare the data of the D-Wave and the observables of the thermal equilibrium of $H_0$ at the obtained effective temperature as shown in Fig.~\ref{fig:app_c_DW}. For $h_0 = 0.2$ and for $t_a>10$, where $s^* > 0.5$, the ground state probability $P_{\mathrm{GS}}$ and the constraint satisfaction probability $P_{\mathrm{const}}$ are close to the value of the thermal equilibrium state, implying that the hypothesis of quasi-static time evolution is valid. On the other hand, in the region where $s^*<0.5$ for all $h_0$ values, $P_{\mathrm{GS}}$ of  the D-Wave  is larger than that of the thermal equilibrium. Furthermore, $P_{\mathrm{const}}$ of the D-Wave is smaller than that of  thermal equilibrium. This result suggests that in the case of $s^*<0.5$, the hypothesis of quasi-static time evolution may not hold, or that the final state of the annealing is affected by the transverse field. 

A similar analysis of the effective temperature is performed for the results of TEBD, as shown in Fig. \ref{fig:app_c_TEBD}. As in the case of  the D-Wave, $P_{\mathrm{GS}}$ is larger, and $P_{\mathrm{const}}$ is smaller than the value of the corresponding thermal equilibrium state. This result suggests that it may be generally observed that the single kink state probability is larger, and the ground state probability is smaller, in non-equilibrium states compared to the thermal equilibrium states at the corresponding effective temperature.

In summary, the annealing time dependence of the effective temperature varies significantly depending on $h_0$. For $h_0=0.2$, the effective temperature decreases toward the hardware temperature while it saturates at higher temperatures for $h_0=1.0$ and $h_0=3.0$. Moreover, the value of effective temperature changes non-monotonically with $h_0$. The mechanism of this nontrivial dependence of the effective temperature on $h_0$ remains to be clarified. In the region where the freeze-out point is small and not in thermal equilibrium of $H_0$, the behaviors of $P_{\mathrm{GS}}$ and $P_{\mathrm{const}}$ exhibit similar characteristics to coherent QA, although the underlying mechanism remains unclear. For a deeper understanding of the results, analysis taking the effects of thermal noise into account is necessary.

\bibliography{main_v4.bib}
\end{document}